\RequirePackage{fixltx2e}
\documentclass[aip,jcp,reprint,floatfix]{revtex4-1}
\usepackage{graphicx} 
\usepackage{color}
\usepackage{dcolumn}
\usepackage{amsmath,amsfonts} %
\usepackage{braket} %
\usepackage[acronym]{glossaries} %
\usepackage[breaklinks, %
colorlinks, %
linkcolor=blue, %
citecolor=blue, %
urlcolor=blue]{hyperref}

\newcommand{\eq}[1]{Eq.~(\ref{#1})} %
\newcommand{\eqs}[1]{Eqs.~(\ref{#1})} %
\newcommand{\fig}[1]{Fig.~\ref{#1}} %
\newcommand{\bea}{\begin{eqnarray}}
\newcommand{\eea}{\end{eqnarray}}
\newcommand{\mat}[1]{\ensuremath{\boldsymbol{#1}}}
\newcommand{\vect}[1]{\ensuremath{\vec{#1} }}
\newcommand{\op}[1]{\ensuremath{\hat{#1}}}
\renewcommand{\ket}[1]{\ensuremath{\left|#1\right\rangle}}
\renewcommand{\bra}[1]{\ensuremath{\left\langle #1\right|}}
\renewcommand{\braket}[2]{\ensuremath{\left\langle #1|#2\right\rangle}}
\newcommand{\braOket}[3]{\ensuremath{\left\langle #1\left|#2\right|#3\right\rangle}}
\renewcommand{\Re}{\operatorname{Re}}
\renewcommand{\Im}{\operatorname{Im}}
\newcommand{\Dim}{\mathcal{D}}
\renewcommand{\d}[1]{\ensuremath{\, \operatorname{d}\!{#1}}}
\newcommand{\e}[1]{\ensuremath{\, \mathrm{e}^{#1}}}

\newacronym{CS}{CS}{coherent state} %
\newacronym{MCTDH}{MCTDH}{multi configuration time-dependent Hartree} %
\newacronym{MCA}{MCA}{moving crude adiabatic} %
\newacronym{AP}{AP}{action time-dependent variational principle} %
\newacronym{TDVP}{TDVP}{time-dependent variational principle} %
\newacronym{MP}{MP}{MacLachlan variational principle} %
\newacronym{EOM}{EOM}{equations of motion} %
\newacronym{AIMS}{AIMS}{ab initio multiple spawning} %
\newacronym{vMCG}{vMCG}{variational multi configuration Gaussian} %
\newacronym{DOF}{DOF}{degrees of freedom} %
\newacronym{CI}{CI}{conical intersection} %
\newacronym{GP}{GP}{geometric phase} %
\newacronym{END}{END}{electron-nuclear dynamics} %
\newacronym{OFQD}{OFQD}{on-the-fly quantum dynamics} %

\begin{document}

\title{Variational nonadiabatic dynamics in the moving crude adiabatic representation: Further merging
of nuclear dynamics and electronic structure}

\author{Lo{\"i}c Joubert-Doriol} \affiliation{Department of Physical
  and Environmental Sciences, University of Toronto Scarborough,
  Toronto, Ontario, M1C 1A4, Canada; and Chemical Physics Theory
  Group, Department of Chemistry, University of Toronto, Toronto,
  Ontario, M5S 3H6, Canada}

\author{Artur F. Izmaylov} %
\affiliation{Department of Physical and Environmental Sciences,
  University of Toronto Scarborough, Toronto, Ontario, M1C 1A4,
  Canada; and Chemical Physics Theory Group, Department of Chemistry,
  University of Toronto, Toronto, Ontario, M5S 3H6, Canada}

\date{\today}

\begin{abstract}
A new methodology of simulating nonadiabatic dynamics using frozen-width Gaussian wavepackets 
within the moving crude adiabatic representation with the on-the-fly evaluation of electronic structure is presented.  
The main feature of the new approach is elimination of any 
global or local model representation of electronic potential energy surfaces, instead, 
the electron-nuclear interaction is treated explicitly using the Gaussian integration. 
As a result, the new scheme does not introduce any uncontrolled approximations.
The employed variational principle ensures the energy conservation and leaves the number 
of electronic and nuclear basis functions as the only parameter 
determining the accuracy. To assess performance of the approach, 
a model with two electronic and two nuclear spacial degrees of freedom
containing conical intersections between potential energy surfaces has been considered. 
Dynamical features associated with nonadiabatic transitions and 
nontrivial geometric (or Berry) phases were successfully reproduced within a limited basis expansion. 
\end{abstract}

\glsresetall
\maketitle

\section{Introduction}
\label{sec:introduction}

One of the popular approaches for on-the-fly simulations of quantum 
nonadiabatic dynamics involves representing the total molecular wavepacket as 
a Born-Huang expansion~\cite{Yang:2009ja,Saita:2012/jcp/22A506}
\bea\label{eq:usual_Psi}
\ket{\Psi(\mat R,t)} & = & \sum_{s=1}^{N_\mathrm{s}}\bigg[\sum_{k=1}^{N_\mathrm{g}} C_{ks}(t) g_k(\mat R,t)\bigg] \ket{\phi_s (\mat R)}
\eea
using a linear combination of $N_\mathrm{g}$ moving
frozen-width Gaussians $g_k(\mat R,t)$ multiplied by $N_\mathrm{s}$ 
adiabatic electronic states $\ket{\phi_s (\mat R)}$.
Gaussians are commonly used for the on-the-fly quantum dynamics\cite{Martinez2000rev,Worth:2004/FD/307,Richings:2015/jpca/12457,meek:2016d} due to their localized character, 
their introduction can be traced back to works of Heller.~\cite{Heller:1975/jcp/1544,Heller:1981/jcp/2923}
In \eq{eq:usual_Psi},
 $C_{ks}(t)$ are time-dependent coefficients, 
and $\mat R$ are the nuclear coordinates. The states $\ket{\phi_s (\mat R)}$ are eigenstates 
of the electronic Hamiltonian $\op H_e[{\mat R}]$ and come naturally from well-developed 
electronic structure software packages. Unfortunately, commonly encountered 
\glspl{CI} of potential energy surfaces\cite{Yarkony:1998/acr/511,Migani:2004/271} 
produce two serious difficulties for fully 
quantum nonadiabatic methods in the adiabatic representation:
1) divergent nonadiabatic couplings from the nuclear kinetic operator acting on the electronic functions,
\cite{meek:2016c,Saxe87,Thompson85} 
and 2) nontrivial \glspl{GP}.\cite{Mead:1979/jcp/2284,Mead:1992/rmp/51,Wittig:2012/pccp/6409,Ryabinkin:2017/acr/1785}

To avoid these problems one can resort to the diabatic 
representation,~\cite{Charlotte:JPCA/2010,Richings:2015/jpca/12457,meek:2016d} 
however, this would require a diabatization procedure, 
which becomes an additional source of approximations.\cite{Mead:1982vm} 
Recently, we discovered that the problems of the adiabatic representation can be resolved 
without abandoning the direct use of the eigenfunctions of the electronic Hamiltonian.\cite{Joubert:2017/jpcl/452} 
The only required modification is to consider the adiabatic wavefunctions parametrically 
dependent on the center of moving nuclear wave-packets
\bea\label{eq:MCA_Psi}
\ket{\Psi(\mat R,t)} & = & \sum_{s=1}^{N_\mathrm{s}} \sum_{k=1}^{N_\mathrm{g}} C_{ks}(t) 
g_k(\mat R,t) \ket{\phi_s(\mat q_k)},
\eea
 where $\mat q_k$ is the center of the $g_k(\mat R,t)$ Gaussian.
As illustrated on a two-state linear vibronic coupling model containing a CI,\cite{Joubert:2017/jpcl/452} 
due to absence of the nuclear coordinate dependence in the adiabatic electronic wavefunctions, 
both problems of the expansion in \eq{eq:usual_Psi} are resolved: 
1) the nuclear kinetic energy does not produce nonadiabatic 
couplings at all, and 2) the nontrivial GP is acquired naturally by the electronic wavefunctions due to 
their parametric dependence on Gaussian centers.
The states $\ket{\phi_s(\mat q_k)}$ are eigenstates of the electronic Hamiltonian only at $\mat q_k$ nuclear
configuration, hence, formally, they are crude adiabatic states.~\cite{Longuet:1961/as/429,Ballhausen:1972/arpc/15} Since these crude adiabatic states are attached to moving nuclear Gaussians, we
 refer to the expansion in \eq{eq:MCA_Psi} as the \gls{MCA} representation. 
Independently, the same representation has been
suggested by Shalashilin and coworkers under the name of time-dependent diabatic 
representation.\cite{Fernandez:2016/pccp/10028}
Although the MCA representation in \eq{eq:MCA_Psi} uses unentangled products of electronic and nuclear basis functions, electron-nuclear correlation similar to that present in the global adiabatic representation [\eq{eq:usual_Psi}] is built when a linear combination of the MCA products is taken. A significant factor contributing to the convergence of the MCA expansion is that both electronic and nuclear parts of an individual product share the same center, the center of a nuclear Gaussian. Therefore, differences between MCA and adiabatic electronic states that are growing with the distance from the Gaussian center are exponentially suppressed by the nuclear Gaussian decay.

To use the MCA representation with electronic structure methods one needs to address 
challenges related to evaluation of new matrix elements originating from non-orthogonality 
of electronic wavefunctions centered at different Gaussians ($\braket{\phi_s(\mat q_k)}{\phi_{s'}(\mat q_l)}\ne \delta_{ss'}$ if $\mat q_k \ne \mat q_l$) and from their non-eigenfunction 
character for the electronic Hamiltonian taken at an arbitrary nuclear point, $H_e[\mat R]$.    
In Refs.~\onlinecite{Makhov:2014/jcp/054110,Fernandez:2016/pccp/10028} 
evaluation of these new matrix elements was done using Taylor series expansions 
around the Gaussian centers. Although making implementation of the formalism feasible, such expansions introduce 
uncontrolled approximations whose quality depends on how strong is the nuclear dependence in
solutions of the electronic problem, electronic wavefunctions and potential energy surfaces (PESs).

In the current work we show that the MCA representation can be used without introducing PESs and their local 
or global approximations. In this exact version, eMCA, the electronic states are calculated on-the-fly 
by solving the electronic problem first, and all the total Hamiltonian matrix elements in the 
\gls{MCA} basis are then calculated exactly along molecular dynamics. 
These exact calculations are possible because the electronic states in eMCA do not 
depend on nuclear coordinates in contrast to other representations (adiabatic or quasi-diabatic)
where saddle point or local harmonic approximations are commonly 
used to describe nuclear coordinate dependence.~\cite{Martinez2000rev,Richings:2017/cpl/606}
eMCA assessment is done on a 2-dimensional 
generalization of the Shin and Metiu model,~\cite{Min:2014/prl/263004} which contains \glspl{CI} 
and exhibits coupled electron-nuclei dynamics. This model contains an explicit 
electron coordinate and therefore requires solving the electronic problem along with the nuclear dynamics
in contrast to vibronic coupling models where electronic \gls{DOF} are represented  
by few implicit diabatic electronic states.~\cite{Kouppel:1984/acp/59}

In principle, the frozen-width Gaussians employed 
in \eqs{eq:usual_Psi} and \eqref{eq:MCA_Psi} can be evolved in several different ways: classically, 
using Born-Oppenheimer~\cite{BenNun:2002tx,Yang:2009ja} or 
Ehrenfest trajectories,~\cite{Saita:2012/jcp/22A506,Makhov:2017/cc/200} or according to the \gls{TDVP} in a full quantum fashion.~\cite{Worth:2004/FD/307,Burghardt:2008iz,Worth:2008/MP/2077,Richings:2017/cpl/606}
We choose to apply the \gls{TDVP} because its variational character accelerates the convergence of 
results with the number of basis functions.~\cite{Worth:2008/MP/2077}
Moreover, the energy is conserved by construction during the dynamics for 
variational \gls{EOM}.~\cite{Kan:1981/pra/2831} In contrast, if classical \gls{EOM} are used,
the energy is conserved only in the complete basis set limit.~\cite{Habershon:2012/jcp/014109}

The paper is organized as follows.
Section~\ref{sec:theory} presents the formalism for the variational full quantum method using the \gls{MCA} representation and discusses the new quantities needed for eMCA.
In Sec.~\ref{sec:results}, we explore the feasibility of eMCA on a realistic system where electronic and nuclear \gls{DOF} are treated explicitly for the on-the-fly dynamics.
Finally, in Sec.~\ref{sec:conclusion} we summarize main results and give future outlook.

\section{Theory}
\label{sec:theory}

\subsection{Time-dependent variational principle for the moving crude adiabatic representation}

Before applying TDVP in the MCA representation we will establish some additional notation 
and few useful relations for the nuclear basis functions expanded as frozen-width Gaussians
\bea\label{eq:cs}
&&\braket{\mat R}{g_k (\mat z_k(t),\mat z_k(t)^*)}  =  \prod_{a=1}^{N_n} \left(\frac{\omega_a}{\pi}\right)^{\frac{\Dim}{4}} G_{ka}(\vect R_{a}) , \\
&& G_{ka}(\vect R_{a})  =  \prod_{\alpha=1}^{\Dim} \e{-\frac{\omega_a}{2}\left[R_{a\alpha}-\sqrt{\frac{2}{\omega_a}}z_{ka\alpha}\right]^2+iz_{ka\alpha}\Im[z_{ka\alpha}]},
\eea
where $\Dim$ is the dimensionality of the space where particles are evolving (3-dimensional for real molecules) and $\{\omega_a\}$ are the width parameters to be chosen for each nucleus.~\cite{Thompson:2010/cp/70}
The Cartesian coordinates are used for nuclear \gls{DOF} in \eq{eq:cs}, this choice was motivated by convenience of 
integrating electron-nuclear interaction terms in the full Hamiltonian.
 The complex parameters $z_{ka\alpha}$ encode the positions
$q_{ka\alpha}(t)=\sqrt{2/\omega_a}\Re[z_{ka\alpha}(t)]$ and the 
momenta $p_{ka\alpha}(t)=\sqrt{2\omega_a}\Im[z_{ka\alpha}(t)]$ of each Gaussian.
These relations stem from a coherent-state form of Gaussians introduced in \eq{eq:cs} 
\bea
\left[\sqrt{\frac{\omega_a}{2}} \op R_{a\alpha} + i \frac{\op P_{a\alpha}}{\sqrt{2\omega_a}} \right] \ket{g_k} & = & z_{ka\alpha} \ket{g_k}, \label{eq:CSalgebra1}
\eea
where $\op P_{a\alpha}$ is the nuclear momentum operator.
Use of coherent states is motivated by their numerical stability in the \gls{EOM} integration.~\cite{Izmaylov:2013/jcp/104115,Saita:2012/jcp/22A506,Makhov:2017/cc/200} 
Throughout this work, time and other parameters will be partially omitted from 
basis and state functions for readability. Also we will use a shorthand notation for the MCA 
electronic states $\ket{\phi_s^k} \equiv \ket{\phi_s(\mat q_k)}$ and for 
the electron-nuclear basis $\ket{\varphi_{ks}} \equiv \ket{g_k}\ket{\phi_s^k}$.

To solve the time-dependent Schr\"odinger equation for the full molecular Hamiltonian, 
$\op H = \op T_n + \op H_e[\mat R]$, we apply the \gls{TDVP},~\cite{Book/Kramer:1981} 
in the least action principle form
\bea
\Im\braket{\delta\Psi}{\dot\Psi+i\op H\Psi} & = & 0,
\eea
where $\ket{\Psi}$ is the molecular wavefunction given by \eq{eq:MCA_Psi}. 
Due to non-analyticity of the \gls{MCA} basis, different forms of the \glspl{TDVP} are not generally equivalent.\cite{Broeckhove:1988wo}
Therefore, to ensure the energy conservation, it is important to apply the least action version 
of the TDVP. 
Then, the \gls{EOM} for the parameters $\{C_{ks},z_{ka\alpha}\}$ become
\bea
i\dot{\mat C} & = & \mat S^{-1}\left[\mat H-i\mat \gamma\right]\mat C, \label{eq:Cdot}\\
\mat B\dot{\mat z} + \mat A\dot{\mat z}^* & = & \mat Y +\overline{\mat Y}, \label{eq:zdot}
\eea
where the involved matrix elements can be written as
\bea
S_{kl,ss'} & = & \braket{\varphi_{ks}}{\varphi_{ls'}}, \label{eq:S}\\
H_{kl,ss'} & = & \braOket{\varphi_{ks}}{\op H}{\varphi_{ls'}}, \label{eq:H}\\
\gamma_{kl,ss'} & = & \braket{\varphi_{ks}}{\dot\varphi_{ls'}}, \label{eq:gamma}
\eea
\bea
Y_{ka\alpha} & = & i\sum_{lss'} C_{ks}^*\braOket{\frac{\partial\varphi_{ks}}{\partial z_{ka\alpha}}}{\op 1-\op{\mathcal{P}}}{\op H\varphi_{ls'}}C_{ls'}, \label{eq:Y}\\
\overline Y_{ka\alpha} & = & i\sum_{lss'} C_{ls'}^*\braOket{\op H\varphi_{ls'}}{\op 1-\op{\mathcal{P}}}{\frac{\partial\varphi_{ks}}{\partial z_{ka\alpha}^*}}C_{ks}, \label{eq:Ybar}\\
A_{kl,ab,\alpha\beta} & = & \sum_{ss'} \Bigg[ C_{ls'}^*\braOket{\frac{\partial\varphi_{ls'}}{\partial z_{lb\beta}}}{\op 1-\op{\mathcal{P}}}{\frac{\partial\varphi_{ks}}{\partial z_{ka\alpha}^*}}C_{ks} \nonumber\\
&&\hspace{0.cm}- C_{ks}^*\braOket{\frac{\partial\varphi_{ks}}{\partial z_{ka\alpha}}}{\op 1-\op{\mathcal{P}}}{\frac{\partial\varphi_{ls'}}{\partial z_{lb\beta}^*}}C_{ls'} \Bigg], \label{eq:A}\\   
B_{kl,ab,\alpha\beta} & = & \sum_{ss'} \Bigg[ C_{ls'}^*\braOket{\frac{\partial\varphi_{ls'}}{\partial z_{lb\beta}^*}}{\op 1-\op{\mathcal{P}}}{\frac{\partial\varphi_{ks}}{\partial z_{ka\alpha}^*}}C_{ks} \nonumber\\
&&\hspace{0.cm}- C_{ks}^*\braOket{\frac{\partial\varphi_{ks}}{\partial z_{ka\alpha}}}{\op 1-\op{\mathcal{P}}}{\frac{\partial\varphi_{ls'}}{\partial z_{lb\beta}}}C_{ls'} \Bigg]. \label{eq:B}   
\eea
Similar equations have been derived for the variational evolution of Gaussian wavepackets in the context of the Gaussian-Multiconfiguration Time-Dependent Hartree (G-MCTDH) and variational Multiconfiguration Gaussian (vMCG) methods.\cite{Burghardt:1999/jcp/2927,Worth:2003/cpl/502,Richings:2015/irpc/269} 
The main source of differences between Eqs. (\ref{eq:Cdot})-(\ref{eq:zdot}) and their vMCG and G-MCTDH counterparts is the use of the parameterization introduced by the MCA representation in Eq.~(\ref{eq:MCA_Psi}). 
Equations~(\ref{eq:Y}-\ref{eq:B}) involve the projector on the non-orthogonal basis 
\bea
\op{\mathcal{P}} & = & \sum_{kl,ss'} \ket{\varphi_{ks}}[\mat S^{-1}]_{kl,ss'}\bra{\varphi_{ls'}}.
\eea
When the basis $\{\ket{\varphi_{ks}}\}$ approaches the complete basis set limit, $\op 1 - \op{\mathcal{P}}$ vanishes and 
eliminates Eq.~(\ref{eq:zdot}) by turning it into the trivial identity, $0=0$. 
This illustrates that there is no need for basis function movement in the complete basis set limit. 

In a more common case of an incomplete basis set, \eq{eq:zdot} can be combined with 
its complex conjugate equivalent and reformulated in a matrix form 
\bea
\begin{pmatrix} \mat Y + \overline{\mat Y} \\ -\mat Y^* - \overline{\mat Y}^*  \end{pmatrix} & = & \begin{pmatrix} \mat B & \mat A \\ \mat A^\dagger & -\mat B^T  \end{pmatrix} \begin{pmatrix} \dot{\mat z} \\ \dot{\mat z}^*  \end{pmatrix}.
\eea
This system of equations is solved as follows
\bea\label{eq:zdotsolve}
\begin{pmatrix} \dot{\mat z} \\ \dot{\mat z}^*  \end{pmatrix} & = & \begin{pmatrix} \mat \beta & \mat \alpha \\ \mat \alpha^\dagger & -\mat \beta^T  \end{pmatrix} \begin{pmatrix} \mat Y + \overline{\mat Y} \\ -\mat Y^* - \overline{\mat Y}^* \end{pmatrix},
\eea
where
\bea
\mat \beta & = & \left[\mat B + \mat A\left(\mat B^T\right)^{-1}\mat A^\dagger\right]^{-1} \\ 
\mat \alpha & = & \mat B^{-1}\mat A\mat \beta^T = \mat \beta\mat A\mat (\mat B^T)^{-1}.
\eea
It is important to note that $\mat A$ and $\mat \alpha$ are antisymmetric, and $\mat B$ and $\mat \beta$ are Hermitian. 

The system energy is conserved by construction, as it can be verified by using \eq{eq:Cdot}, and then 
expressing the energy variation in terms of \eqs{eq:Y}, \eqref{eq:Ybar}, and \eqref{eq:zdotsolve}:
\bea
\dot E & = & 2\Re\braOket{\dot\Psi}{\op H}{\Psi} \nonumber\\
& = & 2\Re\left[ i\overline{\mat Y}^\dagger\dot{\mat z} - i\mat Y^T\dot{\mat z}^* \right] \nonumber\\
& = & 2\Im\left[ \begin{pmatrix} \overline{\mat Y} \\ -\mat Y^* \end{pmatrix}^\dagger \begin{pmatrix} \mat \beta & \mat \alpha \\ \mat \alpha^\dagger & -\mat \beta^T  \end{pmatrix} \begin{pmatrix} \mat Y + \overline{\mat Y} \\ -\mat Y^* - \overline{\mat Y}^* \end{pmatrix} \right], \nonumber\\
& = & -2\Im\left[ \begin{pmatrix} \overline{\mat Y}^* \\ -\mat Y \end{pmatrix}^T \begin{pmatrix} \mat \alpha & \mat \beta \\ -\mat \beta^T & \mat \alpha^\dagger  \end{pmatrix} \begin{pmatrix} \overline{\mat Y}^* \\ -\mat Y \end{pmatrix} \right] = 0.
\eea
In the last two equalities we used that $\mat \beta$ and $\mat \alpha$ are Hermitian and antisymmetric.

As illustrated next, all integrals defined in Eqs.~(\ref{eq:S})-(\ref{eq:B}) can be evaluated numerically 
exactly for the molecular Hamiltonian $\op H$, 
and thus, the current formalism propagates Eqs.~(\ref{eq:Cdot}) and (\ref{eq:zdot}) 
numerically exactly within a finite basis set.


\subsection{Matrix elements}

The matrix elements in \eq{eq:S} involves the product of the nuclear Gaussian overlap with overlap between electronic states obtained at different Gaussian centers
\bea
S_{kl,ss'} & = & \braket{g_{k}}{g_{l}}\braket{\phi_{s}^k}{\phi_{s'}^l} \nonumber\\
 & = & \exp\left[\mat z_k^\dagger\mat z_l-\frac{|\mat z_k|^2+|\mat z_l|^2}{2}\right]\braket{\phi_{s}^k}{\phi_{s'}^l}.
\eea
While the nuclear part has a simple analytic expression, the electronic part requires 
evaluating overlaps between states employing different primitive bases.
Such electronic overlaps appear in other molecular dynamics methods, 
and therefore, have been already efficiently implemented.~\cite{Plasser:2016/jctc/1207}

To treat the Hamiltonian integrals in \eq{eq:H}, first, we added to and subtracted from the Hamiltonian
the electron-nuclei ($\op{V}_\mathrm{en}$) and nuclei-nuclei 
($\op{V}_\mathrm{nn}$) Coulomb terms evaluated at the center of a Gaussian so that we can 
assemble the electronic Hamiltonian at the Gaussian center
\bea
\op H & = & \op{T}_\mathrm{n} + \op H_e [{\mat R}] +(\op V_{en}[{\mat q}]-\op V_{en}[{\mat q}]) +(\op V_{nn}[{\mat q}]-\op V_{nn}[{\mat q}]) \nonumber\\ 
& = & \op{T}_\mathrm{n} + \op H_e [{\mat q}] +(\op V_{en}[{\mat R}]-\op V_{en}[{\mat q}]) +(\op V_{nn}[{\mat R}]-\op V_{nn}[{\mat q}]). \nonumber\\
\eea  
This allows us to reformulate the Hamiltonian integrals as 
\bea\label{eq:H_expanded}
H_{kl,ss'} & = & \braket{\phi_{s}^k}{\phi_{s'}^l}\braOket{g_{k}}{\op{T}_\mathrm{n}+\op{V}_\mathrm{nn}[{\mat R}]}{g_{l}} \nonumber\\
&& +\braOket{\varphi_{ks}}{\op{V}_\mathrm{en}[{\mat R}]-\op{V}_\mathrm{en}[{\mat q}_k]}{\varphi_{ls'}} \nonumber\\
&& +\braket{\varphi_{ks}}{\varphi_{ls'}}\left[\epsilon_s(\mat q_k)-\op {V}_\mathrm{nn}[{\mat q}_k]\right],
\eea
where $\epsilon_s(\mat q_k)$ are the electronic energies at the 
point $\mat q_k$, $\op H_e[\mat q_k] \ket{\phi_{s}^k} = \epsilon_s(\mat q_k) \ket{\phi_{s}^k}$.
The first term on the right-hand side of \eq{eq:H_expanded} can be easily calculated using 
Gaussian integration of the nuclear basis and overlap of the electronic states.
The last term also requires the overlap of the basis functions as well as quantities that are known 
from electronic structure calculations.
In contrast, the second term requires integration of $\op{V}_\mathrm{en}$ 
over electronic states at different nuclear geometries.
To evaluate it, we rewrite the second term as
\bea\label{eq:Ven}
\sum_{a,b=1}^{N_e,N_n}\braOket{\varphi_{ks}}{\frac{Z_b}{|{\vect r}_a-{\vect R}_b|}-\frac{Z_b}{|{\vect r}_a-\vect q_{kb}|}}{\varphi_{ls'}} \nonumber\\
= \int\displaylimits_{-\infty}^{\infty}\hspace{-0.1cm}\d{\vect r}\, \rho_{ss'}^{kl}(\vect r\,) \bar V_{kl}(\vect r\,),
\eea
where $\{{\vect r}_a\}$ are the electronic positions, $Z_b$ are the nuclear charges, 
$\bar V_{kl}(\vect r\,)$ is the electron-nuclei potential ``dressed'' by the Gaussian nuclear functions,
\bea\label{eq:dressedVen}
\bar V_{kl}(\vect r\,) & = & \braket{g_k}{g_l} \sum_{b} \frac{2 Z_b}{\sqrt{\pi}}\Bigg\{\int\displaylimits_{0}^{\sqrt{\omega_b}}\hspace{-0.1cm}\d{u}\e{-u^2\sum\limits_{\alpha}^\Dim\left[{r}_\alpha-\frac{z_{kb\alpha}^*+z_{lb\alpha}}{\sqrt{2\omega_b}}\right]^2} \nonumber\\
 &&\hspace{2.5cm}- \int\displaylimits_{0}^{\infty}\hspace{-0.1cm}\d{u}\e{-u^2\sum\limits_{\alpha}^\Dim\left[{r}_\alpha-q_{kb\alpha}\right]^2}\Bigg\},
\eea
and $\rho_{ss'}^{kl}(\vect r\,)$ is a ``2-point'' electronic transition density
\bea\label{eq:transden}
\rho_{ss'}^{kl}(\vect r\,) & = & \sum_{ij}\chi_i^k(\vect r\,)^*\chi_j^l(\vect r\,) \braOket{\phi_{s}^k}{\op f_i^{k\dagger}\op f_j^l}{\phi_{s'}^l}.
\eea
Here, $\op f_i^{k\dagger}$ and $\op f_i^k$ are the creation and annihilation operators for the $i^\text{th}$ 
molecular orbital, $\chi_i^k(\vect r\,)$, used in the construction of the electronic states at $\mat q_k$ 
(note that orbitals evaluated at $\mat q_k$ and $\mat q_l$ are not orthogonal with respect to each other).
Transition densities between electronic states at different Gaussian centers require expansions in different 
primitive bases, which are obtained using nonunitary orbital transformations.~\cite{Malmqvist:1986/ijqc/479}

The integrals involved in \eq{eq:gamma} can be expanded using the chain rule:
\bea
\gamma_{kl,ss'} & = & \braket{\phi_{s}^k}{\phi_{s'}^l}\left[\braket{g_{k}}{\frac{\partial g_l}{\partial z_{la\alpha}}}\dot z_{la\alpha}+\braket{g_{k}}{\frac{\partial g_l}{\partial z_{la\alpha}^*}}\dot z_{la\alpha}^*\right] \nonumber\\
&&\hspace{2cm} +  \braket{g_{k}}{g_{l}}\braket{\phi_{s}^k}{\frac{\partial \phi_{s'}^l}{\partial q_{la\alpha}}} \dot q_{la\alpha}.
\eea
While the first two terms on the right-hand side can be calculated using coherent state properties and electronic overlaps, the last term involves a quantity that resembles the nonadiabatic couplings for MCA electronic states, $\braket{\phi_{s}^k}{{\partial \phi_{s'}^l}/{\partial q_{la\alpha}}}$.
Generally the number of electronic states considered in simulations can be too small 
to replace these terms by the expansion
\bea\label{eq:2p_NACs}
\braket{\phi_{s}^k}{\frac{\partial \phi_{s'}^l}{\partial q_{la\alpha}}} & = & \sum_u \braket{\phi_{s}^k}{\phi_{u}^l} \braket{\phi_{u}^l}{\frac{\partial \phi_{s'}^l}{\partial q_{la\alpha}}}
\eea
assuming the completeness of the electronic basis set.
The exact calculation of the left hand side of \eq{eq:2p_NACs} requires solving the coupled-perturbed 
equation~\cite{Osamura:1989/tca/113}
\bea\label{eq:CP}
\left( \epsilon_s[\mat q_k] - \op H_e[\mat q_k]  \right) \hspace{-0.1cm}\ket{\frac{\partial \phi_{s}^k}{\partial q_{ka\alpha}}} & = & \left( \frac{\partial\op H_e[\mat q_k]}{\partial q_{ka\alpha}} - \frac{\partial\epsilon_s[\mat q_k]}{\partial q_{ka\alpha}} \right) \hspace{-0.1cm}\ket{\phi_s^k}.\hspace{0.5cm}
\eea
Solving \eq{eq:CP} is usual practice for energy gradients~\cite{Handy:1984/jcp/5031} or derivative couplings~\cite{Lengsfield:1984/jcp/4549}
by projecting analogues of \eq{eq:CP} onto the electronic basis. A similar projection 
technique with the MCA electronic basis was used in the current work for solving
\eq{eq:CP}.

Matrix elements in \eq{eq:Y} to \eq{eq:B} contain integrals, 
\bea\label{eq:extras}
\braket{\frac{\partial \phi_{s}^k}{\partial q_{ka\alpha}}}{\frac{\partial \phi_{s'}^l}{\partial q_{lb\beta}}},\braOket{g_k\frac{\partial \phi_{s}^k}{\partial q_{ka\alpha}}}{\op H}{g_l \phi_{s'}^l}, \nonumber\\
\mathrm{and }\braOket{g_k\phi_{s}^k}{\frac{\partial \op H_e[{\mat R}]}{\partial R_{a\alpha}}}{g_l\phi_{s'}^l},
\eea
which are implemented using components obtained earlier in this section: 
the derivatives of electronic wavefunctions are obtained by solving \eq{eq:CP}, and the matrix elements of 
the differentiated electronic Hamiltonian are evaluated similarly to those in \eq{eq:H_expanded}. 
Note that these integrals are required for the \gls{EOM} obtained employing fully quantum consideration, 
if the basis set dynamic is replaced by classical~\cite{BenNun:2002tx} or 
Ehrenfest dynamics~\cite{Makhov:2017/cc/200} 
these integrals do not appear. 

Thus, the new quantities for which calculations are not already available in electronic structure calculation packages are: the ``2-point'' electronic transition densities given in \eq{eq:transden}, 
and the ``2-point'' electronic states overlap derivatives appearing in \eq{eq:2p_NACs}.
These two quantities are also the most computationally intense parts of the current approach.
They appear in the integrals \eq{eq:H} and \eq{eq:gamma}, whose number scale quadratically with the number of basis functions.
However, the nuclear functions' overlap, which is an exponentially decaying function with respect to differences between the Gaussian parameters, appears in both integrals and can be used for efficient screening~\cite{Gill:1994/cpl/65} to reduce the scaling to linear.

\subsection{Adiabatic nuclear densities} 
\label{sec:addens}

While the \gls{MCA} representation aims to avoid constructing the global adiabatic representation 
during the simulations, one may still want to analyze the results in terms of quantities projected 
onto the global adiabatic representation. This can be done straightforwardly if the projector 
onto an adiabatic state, $\op{\mathcal{Q}}_n[\mat R]= \ket{\phi_n(\mat R)}\bra{\phi_n(\mat R)}$, 
is available. Here, we describe a construction of 
an approximate projection to a $n^\mathrm{th}$ adiabatic state. As a quantity of interest we consider the 
adiabatic nuclear density 
\bea\label{eq:exact_rho}
\rho_n(\mat R) & = & \braOket{\Psi(\mat R)}{\op{\mathcal{Q}}_n[\mat R]}{\Psi(\mat R)}.
\eea
One obvious approximation of $\op{\mathcal{Q}}_n[\mat R]$ can be its first-order Taylor series expansion around 
a particular Gaussian center
\bea\label{eq:Padapp}
\op{\mathcal{Q}}_n[\mat R]& \approx & \op Q^{(1)}_n|_{\mat q_k} \nonumber\\
&=& \op{\mathcal{Q}}_n[\mat q_k] + \sum_{a\alpha} \frac{\partial \op{\mathcal{Q}}_n}{\partial R_{a\alpha}}\bigg|_{\mat q_k}(R_{a\alpha} - q_{ka\alpha}). 
\eea 
However, the choice of the expansion center can be nontrivial considering that $\ket{\Psi(\mat R)}$ is 
expanded using a linear combination of Gaussians located in different places. A special care is required 
in the case of cross-terms, where differently centered Gaussians are originating from 
$\bra{\Psi(\mat R)}$ and $\ket{\Psi(\mat R)}$. One may suggest a double-centered expansion for the 
$\op{\mathcal{Q}}_n[\mat R]$. It turns out that the double-centered expansion does not only 
violate the idempotency but also can introduce spurious double-valuedness in cases with CIs. 
In what follows we adhere to a particular choice that on the one hand provides accurate 
expansion tailored to individual terms in  $\bra{\Psi(\mat R)}$ and $\ket{\Psi(\mat R)}$, and on the 
other hand conserves the correct topological properties of $\rho_n(\mat R)$ associated with GP. 
Using the idempotency of the projector operator we rewrite the density as
\bea
\rho_n(\mat R) & = &\braOket{\Psi(\mat R)}{\op{\mathcal{Q}}_n[\mat R]\op{\mathcal{Q}}_n[\mat R]}{\Psi(\mat R)}\\
 & = & \bigg|\bigg|\sum_{ks}\op{\mathcal{Q}}_n[\mat R]\ket{\phi_s^k}g_k(\mat R)C_{ks}\bigg|\bigg|^2.
\eea
Then, each $\op{\mathcal{Q}}_n[\mat R]\ket{\phi_s^k}$ term is substituted by the first-order 
approximation centered at the Gaussian center $\mat q_k$, which gives
\bea\label{eq:rhoapp}
\rho_n(\mat R) & \approx & \sum_{kl,ss'}C_{ks}^*C_{ls'} g_k(\mat R)^*g_l(\mat R) \nonumber\\
&&\times\braOket{\phi_s^k}{\op Q^{(1)}_n|_{\mat q_k}\op Q^{(1)}_n|_{\mat q_l}}{\phi_{s'}^l}.
\eea
This approach has two more advantages: $\rho_n(\mat R)$ is positively definite and 
can be improved systematically by adding higher order terms in the Taylor expansion of \eq{eq:Padapp}. 
The adiabatic population $P_n^a$ is then calculated by integrating over the nuclear \gls{DOF}
\bea\label{eq:approx3_Pan}
P_n^a & \approx & \int_{-\infty}^{\infty} \d{\mat R}\,\rho_n(\mat R). \nonumber\\
\eea
Since $\rho_n(\mat R)$ and $P_n^a$ are approximated quantities, they do not add up to unity.
To remedy this deficiency, both quantities are renormalized by $\sum_n P_n^a$.

\subsection{Model}
\label{sec:model}

\begin{figure}
  \centering
  \includegraphics[width=0.5\textwidth]{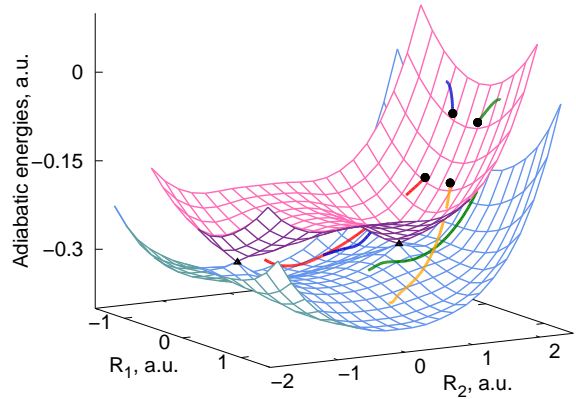}
  \caption{Adiabatic potential energy surfaces for D$_1$ (in blue) and D$_2$ (in pink). Two symmetry allowed \glspl{CI} are represented by the black triangles with coordinates $(0.0,\pm1.2)$. The colored lines are trajectories of the Gaussian basis for the simulation with 4 Gaussian functions during the first $9$ a.u. of time propagation for which black circles are initial positions.}
  \label{fig:CI}
\end{figure}

For numerical illustrations we use a two-dimensional ($\Dim=2$) generalization~\cite{Min:2014/prl/263004} 
of the model Hamiltonian introduced by Shin and Metiu.~\cite{Shin:1995/jcp/9285,Shin:1996/jpc/7867}
This model contains three nuclei and one electron, the positions of two nuclei are fixed, which leaves
one electron ($\vect r$) and one nuclear ($\vect R$) 2D coordinates to consider.
The masses and the charges of the nuclei are $M=10$ and $Z=1$, respectively. 
The Coulomb potential is replaced by a soft Coulomb potential, and the electronic Hamiltonian is
\bea\label{eq:HSM}
\op H_e [{\vect R}] & = & \sum_{\alpha} \frac{\op p_\alpha^2}{2} + V\left(0.5;|{\vect r}-{\vect R}|\right) + V\left(0.5;|{\vect r}-\vect{R}_{+}|\right) \nonumber\\
&&\hspace{-0.2cm} + V\left(0.5;|{\vect r}-\vect{R}_{-}|\right) + V\left(10;|{\vect R}-\vect{R}_{+}|\right) \nonumber\\
&&\hspace{-0.2cm} + V\left(10;|{\vect R}-\vect{R}_{-}|\right) + V\left(10;L\right) + \left(\frac{|{\vect R}|}{3.5}\right)^4, \hspace{0.5cm}
\eea 
where $V(\Delta;x) = (\Delta+x^2)^{-1/2}$, $L=4\sqrt{3}/5$ a.u., $\vect R_{\pm}=(\pm L/2, 0)$ 
are the positions of the fixed protons, and the last term is the two-dimensional quartic potential 
to ensure the system is bounded.
This electronic Hamiltonian gives rise to \glspl{CI} (see \fig{fig:CI}) between the first (D$_1$) and 
second (D$_2$) adiabatic excited states.~\cite{Min:2014/prl/263004,Hader:2017/jcp/074304}
Thus, this model represents a realistic test case for nonadiabatic simulations 
where on-the-fly quantum dynamics can be done exactly.

\subsection{Numerical details}
\label{sec:details}

The eMCA method is tested by modeling nonadiabatic dynamics 
of a wavepacket prepared on the D$_2$ electronic state 
\bea
\ket{\Psi(t=0,\mat R)}&\propto&\e{-\frac{5}{2}|\mat R - \mat q|^2}\ket{\phi_2(\mat q)}.
\eea
where $\mat q = (0,2)$. 
Since electronic parts of \gls{MCA} states match adiabatic electronic states only at the nuclear 
Gaussian center, projection of the initial wavefunction into the adiabatic representation 
produces nonzero $D_1$ state population [see Fig.~\ref{fig:adiab} and ~\ref{fig:DM}-(a)]. 
This initial position of the wavepacket is close to the CI between D$_1$ and D$_2$ 
electronic surfaces. To adequately represent dynamic of the wavepacket and to avoid numerical complications 
associated with the Gaussian center collision with the CI, the initial Gaussian is presented as a linear combination 
of four Gaussians with a smaller width corresponding to $\omega=6.818$. 
The initial parameters for these Gaussians were chosen 
as $\mat C_{k3}=0.287$ and $\mat C_{ks\ne3}=0$, initial positions $\mat q_k=(\pm0.214, 2\pm0.205)$, 
and zero initial momenta.
The trajectories resulting from these initial conditions are shown in \fig{fig:CI} 
and skirt the \gls{CI} in a symmetric manner.

The electronic states are expanded in a direct product basis of harmonic oscillator eigenfunctions.
This harmonic basis is centered at the electronic coordinate origin $(0,0)$ and is defined by its frequency, $0.327$, 
chosen to be the same for both electronic dimensions. The number of the basis functions, maximum quanta, was 
also chosen the same for both dimensions, $n_{x,\mathrm{max}}=n_{y,\mathrm{max}}=n_\mathrm{max}=30$.
The total direct product basis containing 900 states has been pruned to 465 products for which 
$n_{x}+n_{y}\le n_\mathrm{max}$.

Matrices $\mat S$, $\mat B$, and $\mat B+\mat A(\mat B^T)^{-1}\mat A^\dagger$ 
need to be inverted in order to solve \eq{eq:Cdot} and \eq{eq:zdotsolve}. 
These matrices can be close to singular due to overcompletness 
of the Gaussian basis for $\mat S$, and due to small populations $C_{ks}^*C_{ls'}$ for $\mat B$ 
and $\mat B+\mat A(\mat B^T)^{-1}\mat A^\dagger$.
To avoid numerical difficulties, we used a regularization of the inversion procedure that replaces singular values $\lambda\rightarrow \lambda+\varepsilon\exp({-\lambda/\varepsilon})$, where $\varepsilon$ is a threshold. 
Since the accuracy of the $\mat S$ inversion is essential for the quantum propagation of \eq{eq:Cdot},
we use a very small threshold $\varepsilon=10^{-6}$ for this step. 
In contrast, solving \eq{eq:zdot} only gives an optimal evolution of the Gaussian basis but does not impact significantly 
the accuracy when a sufficient number of Gaussians is used.
Thus, we use a larger threshold for \eq{eq:zdot}, $\varepsilon \le10^{-3}$.
All \glspl{EOM} have been solved using the $4^{th}$ order \texttt{ode45} integrator 
implemented in the MATLAB program.~\cite{Matlab:2012b}

\section{Results and discussion}
\label{sec:results}

\begin{figure}
  \centering
  \includegraphics[width=0.5\textwidth]{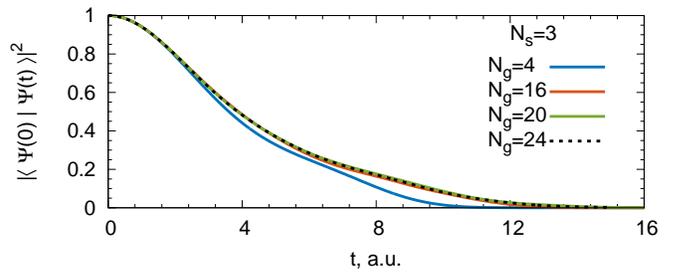}
  \caption{Autocorrelation function norm $|\braket{\Psi(0)}{\Psi(t)}|^2$ for different number of Gaussian functions and $N_s=3$ \gls{MCA} states.}
  \label{fig:auto-Ng}
\end{figure}

\begin{figure}
  \centering
  \includegraphics[width=0.5\textwidth]{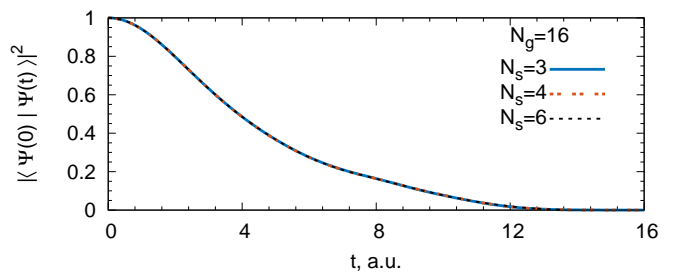}
  \caption{Autocorrelation function norm $|\braket{\Psi(0)}{\Psi(t)}|^2$ for different number of \gls{MCA} states and $N_g=16$ Gaussian functions. All lines are almost indistinguishable.}
  \label{fig:auto-Ns}
\end{figure}

First, we illustrate the convergence with respect to the basis size by comparing 
the norm of the autocorrelation function $|\braket{\Psi(0)}{\Psi(t)}|^2$ for different numbers of Gaussian 
functions in \fig{fig:auto-Ng} and electronic states in \fig{fig:auto-Ns}.
The convergence with respect to the latter is already achieved for $N_\mathrm{s}=3$ (see \fig{fig:auto-Ns})
due to a large energy gap between a cluster of the first three states, D$_{0-2}$, and the rest  
(e.g., at the initial geometry, $\epsilon_2-\epsilon_1=0.106$ a.u. and $\epsilon_3-\epsilon_2=0.326$ a.u.).


\begin{figure}
  \centering
  \includegraphics[width=0.5\textwidth]{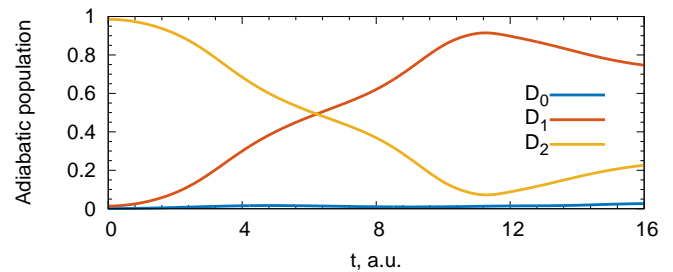}
  \caption{Population of the three first adiabatic states for $N_g=4$ and $N_s=3$.}
  \label{fig:adiab}
\end{figure}

\begin{figure}
  \centering
  \begin{tabular}{cc}
    \raisebox{3.5cm}{(a)}&\includegraphics[width=0.45\textwidth]{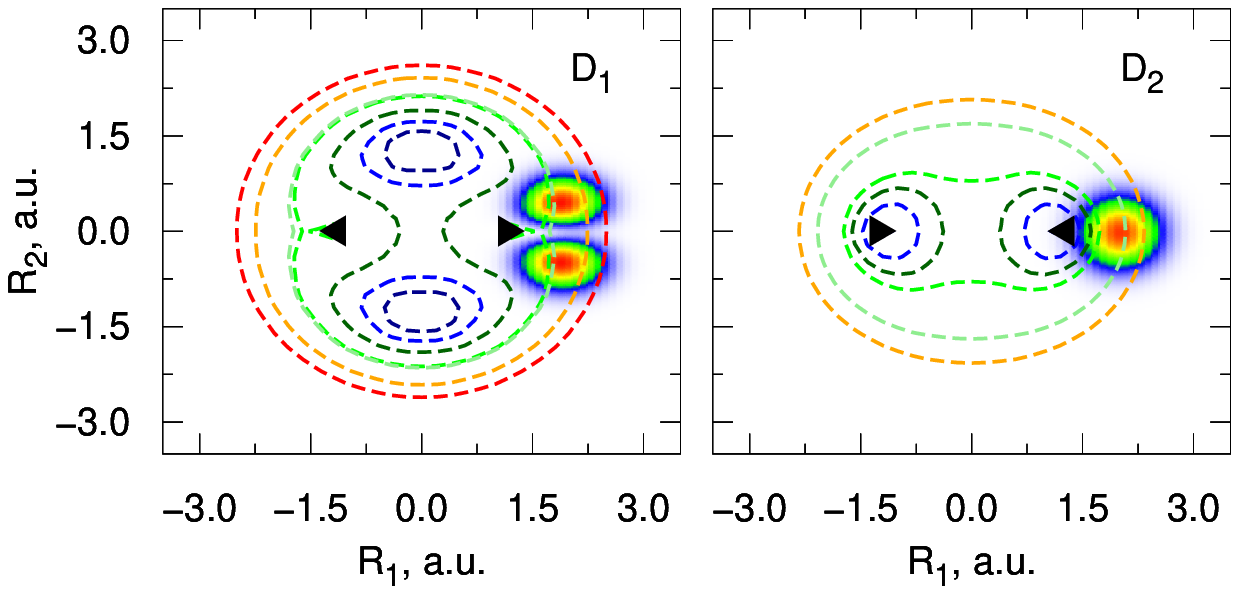}\\
    \raisebox{3.5cm}{(b)}&\includegraphics[width=0.45\textwidth]{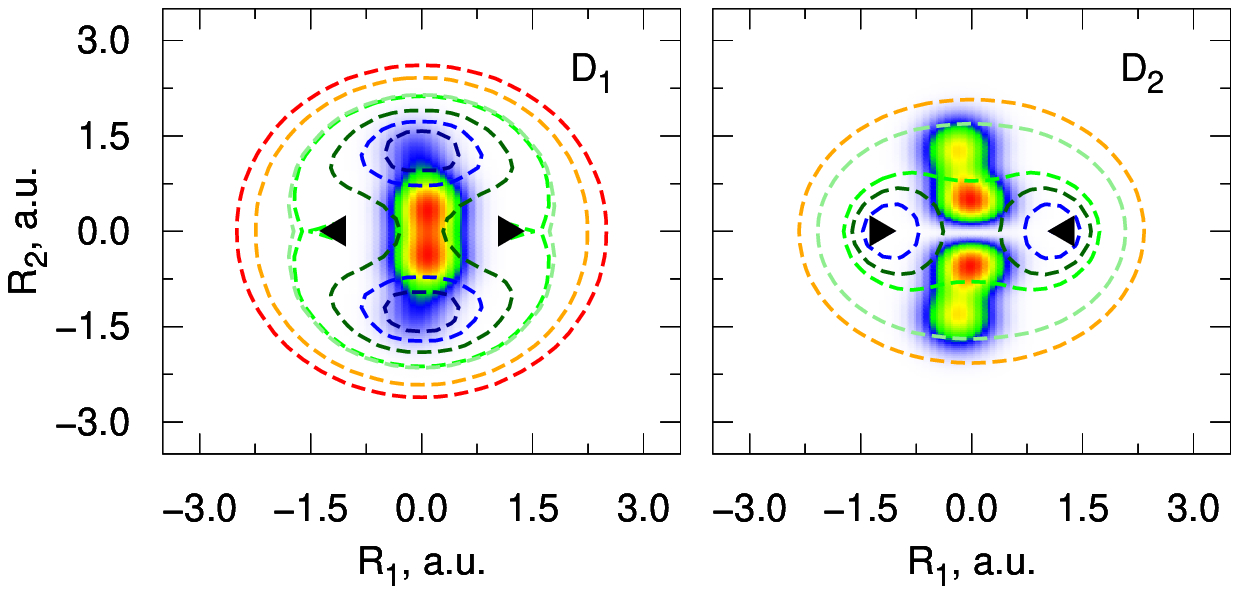}\\
    \raisebox{3.5cm}{(c)}&\includegraphics[width=0.45\textwidth]{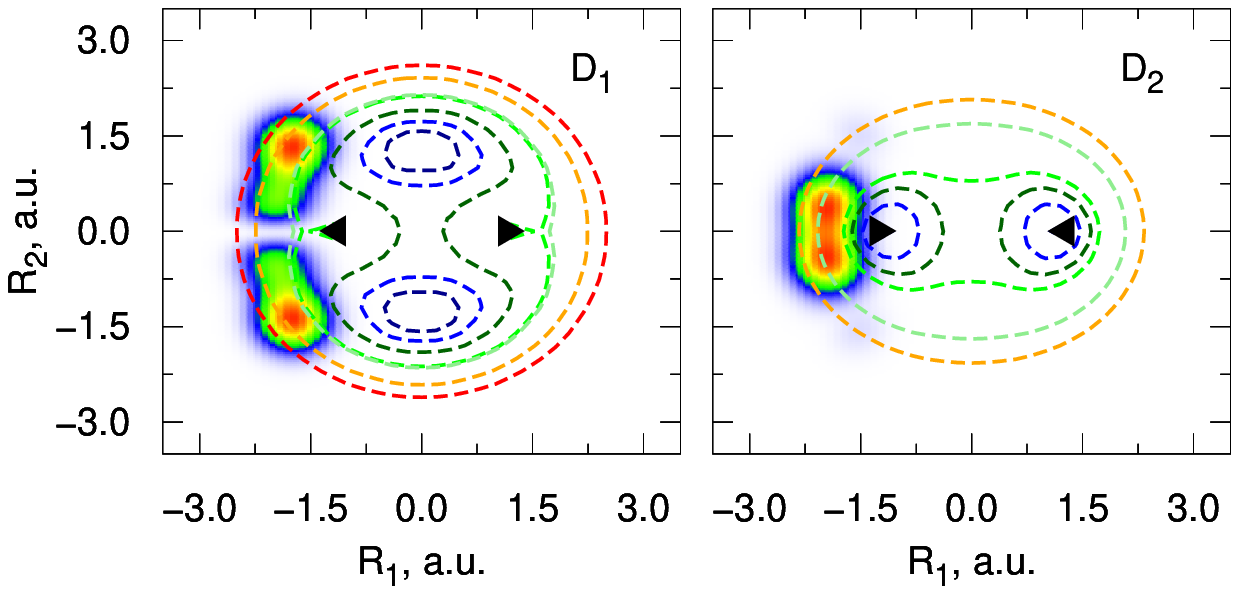}
  \end{tabular}
  \caption{Nuclear density of adiabatic states D$_1$ (on the left) and D$_2$ (on the right) at different time: (a) $t=0$, (b) $t=13.9$, (c) $t=26.2$. Contours of the adiabatic potential energy surfaces of D$_1$ and D$_2$ are superimposed to the densities. Triangles indicate the \glspl{CI} positions.}
  \label{fig:DM}
\end{figure}

Nonadiabatic dynamics is illustrated in \fig{fig:adiab} for the case of $N_g=4$ and $N_s=3$.
One of the main features of this dynamics is radiationless population transfer 
between adiabatic states, which takes place in the vicinity of the \gls{CI}.
Another feature is related to a nontrivial geometric (or Berry) phase induced by \glspl{CI} between D$_1$ and D$_2$, 
the nuclear density corresponding to an adiabatic state exhibits a nodal line 
upon skirting one \gls{CI},~\cite{Schon:1995/jcp/9292,Ryabinkin:2017/acr/1785,Joubert:2017/jpcl/452} 
this node disappears after encircling a second \gls{CI}.~\cite{Zwanziger:1987/jcp/2954,Domcke:2012/arpc/325}
Since the wavepacket starts on D$_2$ it must display a nodal line between the two \glspl{CI} on D$_2$ 
and the absence of the nodal line between the \glspl{CI} on D$_1$.
Indeed, these nodal features can be observed in our simulations with a nodal line appearing on 
D$_1$ for $1.2<R_1<-1.2$ [see \fig{fig:DM}-(a,c)] and on D$_2$ for $-1.2<R_1<1.2$ [see \fig{fig:DM}-(b)].

Figures~\ref{fig:DM}-(a-c) used the approximate expressions for the adiabatic nuclear densities 
derived in \eq{eq:rhoapp}. We have also calculated the corresponding exact 
adiabatic densities on a grid by generating the exact adiabatic states at each point.
The approximate and exact densities were found to be visually indistinguishable which confirms 
the quality of the employed approximation. 

Deviation of the norm of the wavefunction as well as the relative energy deviation 
$\braOket{\Psi}{\op H}{\Psi}/E_0$ ($E_0$ is chosen as the energy difference of 
the adiabatic states D$_1$ and D$_2$ at the initial position) was smaller than $10^{-5}$ 
in all our simulations.
This number is the numerical precision of the current method considering the error 
introduced by the regularization to solve \eq{eq:Cdot} and error accumulation  
along the propagation.

\section{Conclusions}
\label{sec:conclusion}


Using the 2D model with explicit electron and nuclear DOF 
we demonstrated feasibility of the eMCA approach for on-the-fly simulations of nonadiabatic dynamics 
without approximating the involved matrix elements.
Owing to its capability for exact calculation of matrix elements, the eMCA approach 
provides a route to quantum dynamics with controlled approximations.
eMCA is fully variational, which ensures the system energy conservation at any setup. 
The only parameter defining the accuracy of the eMCA approach is the basis set size.
It was shown that the \gls{MCA} expansion has a convergence 
with the number of explicitly included electronic states similar to what would be expected from 
the conventional Born-Huang expansion using the adiabatic electronic states. 
This can be rationalized considering that even though \gls{MCA} involves crude adiabatic states
their interactions due to nonorthogonality in different nuclear geometry points 
are attenuated by exponentially decaying overlaps of attached nuclear Gaussian functions.
Systematic improvement of the MCA representation with respect to the number of nuclear Gaussian functions
can be done using spawning\cite{BenNun:2002tx,Izmaylov:2013/jcp/104115} and cloning\cite{Makhov:2014/jcp/054110,Makhov:2017/cc/200,Izmaylov:2017/jpcl/1793} approaches.

  
Implementing eMCA in conjunction with electronic structure methods 
will require the calculation of the electronic transition densities and electronic state overlap derivatives 
for different nuclear geometries. Electronic structure algorithms for finding these quantities 
with electronic functions at the same nuclear geometry are already available and can be extended for eMCA. 
Furthermore, eMCA can be extended to even larger system using 
quantum-classical treatment~\cite{Romer:2013/mp/3618} or non-unitary dynamics 
combined with the system-environment partitioning.~\cite{Joubert:2015/jcp/134107}

Finally, the explicit treatment of the electronic \gls{DOF} will make eMCA a method of choice for 
studying the electronic dynamics on short timescales of femto- or sub-femtoseconds while treating 
the electron-nuclei interaction exactly. This interaction is essential for elucidating  
a role of the nuclear motion for electronic decoherence.~\cite{Vacher:2017/prl/083001}

\section{Acknowledgements}

The authors thank Ilya Ryabinkin for helpful discussions.
This work was supported by a Sloan Research Fellowship, Natural Sciences and 
Engineering Research Council of Canada (NSERC).


\begin{thebibliography}{57}%
\makeatletter
\providecommand \@ifxundefined [1]{%
 \@ifx{#1\undefined}
}%
\providecommand \@ifnum [1]{%
 \ifnum #1\expandafter \@firstoftwo
 \else \expandafter \@secondoftwo
 \fi
}%
\providecommand \@ifx [1]{%
 \ifx #1\expandafter \@firstoftwo
 \else \expandafter \@secondoftwo
 \fi
}%
\providecommand \natexlab [1]{#1}%
\providecommand \enquote  [1]{``#1''}%
\providecommand \bibnamefont  [1]{#1}%
\providecommand \bibfnamefont [1]{#1}%
\providecommand \citenamefont [1]{#1}%
\providecommand \href@noop [0]{\@secondoftwo}%
\providecommand \href [0]{\begingroup \@sanitize@url \@href}%
\providecommand \@href[1]{\@@startlink{#1}\@@href}%
\providecommand \@@href[1]{\endgroup#1\@@endlink}%
\providecommand \@sanitize@url [0]{\catcode `\\12\catcode `\$12\catcode
  `\&12\catcode `\#12\catcode `\^12\catcode `\_12\catcode `\%12\relax}%
\providecommand \@@startlink[1]{}%
\providecommand \@@endlink[0]{}%
\providecommand \url  [0]{\begingroup\@sanitize@url \@url }%
\providecommand \@url [1]{\endgroup\@href {#1}{\urlprefix }}%
\providecommand \urlprefix  [0]{URL }%
\providecommand \Eprint [0]{\href }%
\providecommand \doibase [0]{http://dx.doi.org/}%
\providecommand \selectlanguage [0]{\@gobble}%
\providecommand \bibinfo  [0]{\@secondoftwo}%
\providecommand \bibfield  [0]{\@secondoftwo}%
\providecommand \translation [1]{[#1]}%
\providecommand \BibitemOpen [0]{}%
\providecommand \bibitemStop [0]{}%
\providecommand \bibitemNoStop [0]{.\EOS\space}%
\providecommand \EOS [0]{\spacefactor3000\relax}%
\providecommand \BibitemShut  [1]{\csname bibitem#1\endcsname}%
\let\auto@bib@innerbib\@empty
\bibitem [{\citenamefont {Yang}\ \emph {et~al.}(2009)\citenamefont {Yang},
  \citenamefont {Coe}, \citenamefont {Kaduk},\ and\ \citenamefont
  {Mart{\'\i}nez}}]{Yang:2009ja}%
  \BibitemOpen
  \bibfield  {author} {\bibinfo {author} {\bibfnamefont {S.}~\bibnamefont
  {Yang}}, \bibinfo {author} {\bibfnamefont {J.~D.}\ \bibnamefont {Coe}},
  \bibinfo {author} {\bibfnamefont {B.}~\bibnamefont {Kaduk}}, \ and\ \bibinfo
  {author} {\bibfnamefont {T.~J.}\ \bibnamefont {Mart{\'\i}nez}},\ }\href
  {\doibase 10.1063/1.3103930} {\bibfield  {journal} {\bibinfo  {journal} {J.
  Chem. Phys.}\ }\textbf {\bibinfo {volume} {130}},\ \bibinfo {pages} {134113}
  (\bibinfo {year} {2009})}\BibitemShut {NoStop}%
\bibitem [{\citenamefont {Saita}\ and\ \citenamefont
  {Shalashilin}(2012)}]{Saita:2012/jcp/22A506}%
  \BibitemOpen
  \bibfield  {author} {\bibinfo {author} {\bibfnamefont {K.}~\bibnamefont
  {Saita}}\ and\ \bibinfo {author} {\bibfnamefont {D.~V.}\ \bibnamefont
  {Shalashilin}},\ }\href {\doibase 10.1063/1.4734313} {\bibfield  {journal}
  {\bibinfo  {journal} {J. Chem. Phys.}\ }\textbf {\bibinfo {volume} {137}},\
  \bibinfo {pages} {22A506} (\bibinfo {year} {2012})}\BibitemShut {NoStop}%
\bibitem [{\citenamefont {Ben-Nun}, \citenamefont {Quenneville},\ and\
  \citenamefont {Martinez}(2000)}]{Martinez2000rev}%
  \BibitemOpen
  \bibfield  {author} {\bibinfo {author} {\bibfnamefont {M.}~\bibnamefont
  {Ben-Nun}}, \bibinfo {author} {\bibfnamefont {J.}~\bibnamefont
  {Quenneville}}, \ and\ \bibinfo {author} {\bibfnamefont {T.}~\bibnamefont
  {Martinez}},\ }\href@noop {} {\bibfield  {journal} {\bibinfo  {journal} {J.
  Phys. Chem. A}\ }\textbf {\bibinfo {volume} {104}},\ \bibinfo {pages} {5161}
  (\bibinfo {year} {2000})}\BibitemShut {NoStop}%
\bibitem [{\citenamefont {Worth}, \citenamefont {Robb},\ and\ \citenamefont
  {Burghardt}(2004)}]{Worth:2004/FD/307}%
  \BibitemOpen
  \bibfield  {author} {\bibinfo {author} {\bibfnamefont {G.~A.}\ \bibnamefont
  {Worth}}, \bibinfo {author} {\bibfnamefont {M.~A.}\ \bibnamefont {Robb}}, \
  and\ \bibinfo {author} {\bibfnamefont {I.}~\bibnamefont {Burghardt}},\ }\href
  {\doibase 10.1039/B314253A} {\bibfield  {journal} {\bibinfo  {journal}
  {Faraday Discuss.}\ }\textbf {\bibinfo {volume} {127}},\ \bibinfo {pages}
  {307} (\bibinfo {year} {2004})}\BibitemShut {NoStop}%
\bibitem [{\citenamefont {Richings}\ and\ \citenamefont
  {Worth}(2015)}]{Richings:2015/jpca/12457}%
  \BibitemOpen
  \bibfield  {author} {\bibinfo {author} {\bibfnamefont {G.~W.}\ \bibnamefont
  {Richings}}\ and\ \bibinfo {author} {\bibfnamefont {G.~A.}\ \bibnamefont
  {Worth}},\ }\href {\doibase 10.1021/acs.jpca.5b07921} {\bibfield  {journal}
  {\bibinfo  {journal} {J. Phys. Chem. A}\ }\textbf {\bibinfo {volume} {119}},\
  \bibinfo {pages} {12457} (\bibinfo {year} {2015})}\BibitemShut {NoStop}%
\bibitem [{\citenamefont {Meek}\ and\ \citenamefont
  {Levine}(2016{\natexlab{a}})}]{meek:2016d}%
  \BibitemOpen
  \bibfield  {author} {\bibinfo {author} {\bibfnamefont {G.~A.}\ \bibnamefont
  {Meek}}\ and\ \bibinfo {author} {\bibfnamefont {B.~G.}\ \bibnamefont
  {Levine}},\ }\href@noop {} {\bibfield  {journal} {\bibinfo  {journal} {J.
  Chem. Phys.}\ }\textbf {\bibinfo {volume} {145}},\ \bibinfo {pages} {184103}
  (\bibinfo {year} {2016}{\natexlab{a}})}\BibitemShut {NoStop}%
\bibitem [{\citenamefont {Heller}(1975)}]{Heller:1975/jcp/1544}%
  \BibitemOpen
  \bibfield  {author} {\bibinfo {author} {\bibfnamefont {E.~J.}\ \bibnamefont
  {Heller}},\ }\href {\doibase 10.1063/1.430620} {\bibfield  {journal}
  {\bibinfo  {journal} {J. Chem. Phys.}\ }\textbf {\bibinfo {volume} {62}},\
  \bibinfo {pages} {1544} (\bibinfo {year} {1975})}\BibitemShut {NoStop}%
\bibitem [{\citenamefont {Heller}(1981)}]{Heller:1981/jcp/2923}%
  \BibitemOpen
  \bibfield  {author} {\bibinfo {author} {\bibfnamefont {E.~J.}\ \bibnamefont
  {Heller}},\ }\href {\doibase 10.1063/1.442382} {\bibfield  {journal}
  {\bibinfo  {journal} {J. Chem. Phys.}\ }\textbf {\bibinfo {volume} {75}},\
  \bibinfo {pages} {2923} (\bibinfo {year} {1981})}\BibitemShut {NoStop}%
\bibitem [{\citenamefont {Yarkony}(1998)}]{Yarkony:1998/acr/511}%
  \BibitemOpen
  \bibfield  {author} {\bibinfo {author} {\bibfnamefont {D.~R.}\ \bibnamefont
  {Yarkony}},\ }\href {\doibase 10.1021/ar970113w} {\bibfield  {journal}
  {\bibinfo  {journal} {Acc. Chem. Res.}\ }\textbf {\bibinfo {volume} {31}},\
  \bibinfo {pages} {511} (\bibinfo {year} {1998})}\BibitemShut {NoStop}%
\bibitem [{\citenamefont {Migani}\ and\ \citenamefont
  {Olivucci}(2004)}]{Migani:2004/271}%
  \BibitemOpen
  \bibfield  {author} {\bibinfo {author} {\bibfnamefont {A.}~\bibnamefont
  {Migani}}\ and\ \bibinfo {author} {\bibfnamefont {M.}~\bibnamefont
  {Olivucci}},\ }in\ \href@noop {} {\emph {\bibinfo {booktitle} {{Conical
  Intersection Electronic Structure, Dynamics and Spectroscopy}}}},\ \bibinfo
  {editor} {edited by\ \bibinfo {editor} {\bibfnamefont {W.}~\bibnamefont
  {Domcke}}, \bibinfo {editor} {\bibfnamefont {D.~R.}\ \bibnamefont {Yarkony}},
  \ and\ \bibinfo {editor} {\bibfnamefont {H.}~\bibnamefont {K\"{o}ppel}}}\
  (\bibinfo  {publisher} {World Scientific},\ \bibinfo {address} {New Jersey},\
  \bibinfo {year} {2004})\ p.\ \bibinfo {pages} {271}\BibitemShut {NoStop}%
\bibitem [{\citenamefont {Meek}\ and\ \citenamefont
  {Levine}(2016{\natexlab{b}})}]{meek:2016c}%
  \BibitemOpen
  \bibfield  {author} {\bibinfo {author} {\bibfnamefont {G.~A.}\ \bibnamefont
  {Meek}}\ and\ \bibinfo {author} {\bibfnamefont {B.~G.}\ \bibnamefont
  {Levine}},\ }\href@noop {} {\bibfield  {journal} {\bibinfo  {journal} {J.
  Chem. Phys.}\ }\textbf {\bibinfo {volume} {144}},\ \bibinfo {pages} {184109}
  (\bibinfo {year} {2016}{\natexlab{b}})}\BibitemShut {NoStop}%
\bibitem [{\citenamefont {Saxe}\ and\ \citenamefont {Yarkony}(1987)}]{Saxe87}%
  \BibitemOpen
  \bibfield  {author} {\bibinfo {author} {\bibfnamefont {P.}~\bibnamefont
  {Saxe}}\ and\ \bibinfo {author} {\bibfnamefont {D.~R.}\ \bibnamefont
  {Yarkony}},\ }\href {\doibase 10.1063/1.452621} {\bibfield  {journal}
  {\bibinfo  {journal} {J. Chem. Phys.}\ }\textbf {\bibinfo {volume} {86}},\
  \bibinfo {pages} {321} (\bibinfo {year} {1987})}\BibitemShut {NoStop}%
\bibitem [{\citenamefont {Thompson}, \citenamefont {Truhlar},\ and\
  \citenamefont {Mead}(1985)}]{Thompson85}%
  \BibitemOpen
  \bibfield  {author} {\bibinfo {author} {\bibfnamefont {T.~C.}\ \bibnamefont
  {Thompson}}, \bibinfo {author} {\bibfnamefont {D.~G.}\ \bibnamefont
  {Truhlar}}, \ and\ \bibinfo {author} {\bibfnamefont {C.~A.}\ \bibnamefont
  {Mead}},\ }\href {\doibase 10.1063/1.448333} {\bibfield  {journal} {\bibinfo
  {journal} {J. Chem. Phys.}\ }\textbf {\bibinfo {volume} {82}},\ \bibinfo
  {pages} {2392} (\bibinfo {year} {1985})}\BibitemShut {NoStop}%
\bibitem [{\citenamefont {Mead}\ and\ \citenamefont
  {Truhlar}(1979)}]{Mead:1979/jcp/2284}%
  \BibitemOpen
  \bibfield  {author} {\bibinfo {author} {\bibfnamefont {C.~A.}\ \bibnamefont
  {Mead}}\ and\ \bibinfo {author} {\bibfnamefont {D.~G.}\ \bibnamefont
  {Truhlar}},\ }\href {\doibase 10.1063/1.437734} {\bibfield  {journal}
  {\bibinfo  {journal} {J. Chem. Phys.}\ }\textbf {\bibinfo {volume} {70}},\
  \bibinfo {pages} {2284} (\bibinfo {year} {1979})}\BibitemShut {NoStop}%
\bibitem [{\citenamefont {Mead}(1992)}]{Mead:1992/rmp/51}%
  \BibitemOpen
  \bibfield  {author} {\bibinfo {author} {\bibfnamefont {C.~A.}\ \bibnamefont
  {Mead}},\ }\href {\doibase 10.1103/RevModPhys.64.51} {\bibfield  {journal}
  {\bibinfo  {journal} {Rev. Mod. Phys.}\ }\textbf {\bibinfo {volume} {64}},\
  \bibinfo {pages} {51} (\bibinfo {year} {1992})}\BibitemShut {NoStop}%
\bibitem [{\citenamefont {Wittig}(2012)}]{Wittig:2012/pccp/6409}%
  \BibitemOpen
  \bibfield  {author} {\bibinfo {author} {\bibfnamefont {C.}~\bibnamefont
  {Wittig}},\ }\href {\doibase 10.1039/c2cp22974a} {\bibfield  {journal}
  {\bibinfo  {journal} {Phys. Chem. Chem. Phys.}\ }\textbf {\bibinfo {volume}
  {14}},\ \bibinfo {pages} {6409} (\bibinfo {year} {2012})}\BibitemShut
  {NoStop}%
\bibitem [{\citenamefont {Ryabinkin}, \citenamefont {Joubert-Doriol},\ and\
  \citenamefont {Izmaylov}(2017)}]{Ryabinkin:2017/acr/1785}%
  \BibitemOpen
  \bibfield  {author} {\bibinfo {author} {\bibfnamefont {I.~G.}\ \bibnamefont
  {Ryabinkin}}, \bibinfo {author} {\bibfnamefont {L.}~\bibnamefont
  {Joubert-Doriol}}, \ and\ \bibinfo {author} {\bibfnamefont {A.~F.}\
  \bibnamefont {Izmaylov}},\ }\href {\doibase 10.1021/acs.accounts.7b00220}
  {\bibfield  {journal} {\bibinfo  {journal} {Acc. Chem. Res.}\ }\textbf
  {\bibinfo {volume} {50}},\ \bibinfo {pages} {1785} (\bibinfo {year}
  {2017})}\BibitemShut {NoStop}%
\bibitem [{\citenamefont {Allan}\ \emph {et~al.}(2010)\citenamefont {Allan},
  \citenamefont {Lasorne}, \citenamefont {Worth},\ and\ \citenamefont
  {Robb}}]{Charlotte:JPCA/2010}%
  \BibitemOpen
  \bibfield  {author} {\bibinfo {author} {\bibfnamefont {C.~S.~M.}\
  \bibnamefont {Allan}}, \bibinfo {author} {\bibfnamefont {B.}~\bibnamefont
  {Lasorne}}, \bibinfo {author} {\bibfnamefont {G.~A.}\ \bibnamefont {Worth}},
  \ and\ \bibinfo {author} {\bibfnamefont {M.~A.}\ \bibnamefont {Robb}},\
  }\href {\doibase 10.1021/jp101574b} {\bibfield  {journal} {\bibinfo
  {journal} {J. Phys. Chem. A}\ }\textbf {\bibinfo {volume} {114}},\ \bibinfo
  {pages} {8713} (\bibinfo {year} {2010})},\ \bibinfo {note} {pMID: 20499843},\
  \Eprint {http://arxiv.org/abs/https://doi.org/10.1021/jp101574b}
  {https://doi.org/10.1021/jp101574b} \BibitemShut {NoStop}%
\bibitem [{\citenamefont {Mead}\ and\ \citenamefont
  {Truhlar}(1982)}]{Mead:1982vm}%
  \BibitemOpen
  \bibfield  {author} {\bibinfo {author} {\bibfnamefont {C.~A.}\ \bibnamefont
  {Mead}}\ and\ \bibinfo {author} {\bibfnamefont {D.~G.}\ \bibnamefont
  {Truhlar}},\ }\href@noop {} {\bibfield  {journal} {\bibinfo  {journal} {J.
  Chem. Phys.}\ }\textbf {\bibinfo {volume} {77}},\ \bibinfo {pages} {6090}
  (\bibinfo {year} {1982})}\BibitemShut {NoStop}%
\bibitem [{\citenamefont {Joubert-Doriol}\ \emph {et~al.}(2017)\citenamefont
  {Joubert-Doriol}, \citenamefont {Sivasubramanium}, \citenamefont
  {Ryabinkin},\ and\ \citenamefont {Izmaylov}}]{Joubert:2017/jpcl/452}%
  \BibitemOpen
  \bibfield  {author} {\bibinfo {author} {\bibfnamefont {L.}~\bibnamefont
  {Joubert-Doriol}}, \bibinfo {author} {\bibfnamefont {J.}~\bibnamefont
  {Sivasubramanium}}, \bibinfo {author} {\bibfnamefont {I.~G.}\ \bibnamefont
  {Ryabinkin}}, \ and\ \bibinfo {author} {\bibfnamefont {A.~F.}\ \bibnamefont
  {Izmaylov}},\ }\href {\doibase 10.1021/acs.jpclett.6b02660} {\bibfield
  {journal} {\bibinfo  {journal} {J. Phys. Chem. Lett.}\ }\textbf {\bibinfo
  {volume} {8}},\ \bibinfo {pages} {452} (\bibinfo {year} {2017})}\BibitemShut
  {NoStop}%
\bibitem [{\citenamefont {Longuet-Higgins}(1961)}]{Longuet:1961/as/429}%
  \BibitemOpen
  \bibfield  {author} {\bibinfo {author} {\bibfnamefont {H.}~\bibnamefont
  {Longuet-Higgins}},\ }in\ \href@noop {} {\emph {\bibinfo {booktitle} {Adv.
  Spectrosc.}}}\ (\bibinfo {year} {1961})\ p.\ \bibinfo {pages}
  {429}\BibitemShut {NoStop}%
\bibitem [{\citenamefont {Ballhausen}\ and\ \citenamefont
  {Hansen}(1972)}]{Ballhausen:1972/arpc/15}%
  \BibitemOpen
  \bibfield  {author} {\bibinfo {author} {\bibfnamefont {C.}~\bibnamefont
  {Ballhausen}}\ and\ \bibinfo {author} {\bibfnamefont {A.~E.}\ \bibnamefont
  {Hansen}},\ }\href@noop {} {\bibfield  {journal} {\bibinfo  {journal} {Ann.
  Rev. Phys. Chem.}\ }\textbf {\bibinfo {volume} {23}},\ \bibinfo {pages} {15}
  (\bibinfo {year} {1972})}\BibitemShut {NoStop}%
\bibitem [{\citenamefont {Fernandez-Alberti}\ \emph {et~al.}(2016)\citenamefont
  {Fernandez-Alberti}, \citenamefont {Makhov}, \citenamefont {Tretiak},\ and\
  \citenamefont {Shalashilin}}]{Fernandez:2016/pccp/10028}%
  \BibitemOpen
  \bibfield  {author} {\bibinfo {author} {\bibfnamefont {S.}~\bibnamefont
  {Fernandez-Alberti}}, \bibinfo {author} {\bibfnamefont {D.~V.}\ \bibnamefont
  {Makhov}}, \bibinfo {author} {\bibfnamefont {S.}~\bibnamefont {Tretiak}}, \
  and\ \bibinfo {author} {\bibfnamefont {D.~V.}\ \bibnamefont {Shalashilin}},\
  }\href {\doibase 10.1039/C5CP07332D} {\bibfield  {journal} {\bibinfo
  {journal} {Phys. Chem. Chem. Phys.}\ }\textbf {\bibinfo {volume} {18}},\
  \bibinfo {pages} {10028} (\bibinfo {year} {2016})}\BibitemShut {NoStop}%
\bibitem [{\citenamefont {Makhov}\ \emph {et~al.}(2014)\citenamefont {Makhov},
  \citenamefont {Glover}, \citenamefont {Martinez},\ and\ \citenamefont
  {Shalashilin}}]{Makhov:2014/jcp/054110}%
  \BibitemOpen
  \bibfield  {author} {\bibinfo {author} {\bibfnamefont {D.~V.}\ \bibnamefont
  {Makhov}}, \bibinfo {author} {\bibfnamefont {W.~J.}\ \bibnamefont {Glover}},
  \bibinfo {author} {\bibfnamefont {T.~J.}\ \bibnamefont {Martinez}}, \ and\
  \bibinfo {author} {\bibfnamefont {D.~V.}\ \bibnamefont {Shalashilin}},\
  }\href {\doibase 10.1063/1.4891530} {\bibfield  {journal} {\bibinfo
  {journal} {J. Chem. Phys.}\ }\textbf {\bibinfo {volume} {141}},\ \bibinfo
  {pages} {054110} (\bibinfo {year} {2014})}\BibitemShut {NoStop}%
\bibitem [{\citenamefont {Richings}\ and\ \citenamefont
  {Worth}(2017)}]{Richings:2017/cpl/606}%
  \BibitemOpen
  \bibfield  {author} {\bibinfo {author} {\bibfnamefont {G.~W.}\ \bibnamefont
  {Richings}}\ and\ \bibinfo {author} {\bibfnamefont {G.~A.}\ \bibnamefont
  {Worth}},\ }\href {\doibase 10.1016/j.cplett.2017.03.032} {\bibfield
  {journal} {\bibinfo  {journal} {Chem. Phys. Lett.}\ }\textbf {\bibinfo
  {volume} {683}},\ \bibinfo {pages} {606} (\bibinfo {year}
  {2017})}\BibitemShut {NoStop}%
\bibitem [{\citenamefont {Min}\ \emph {et~al.}(2014)\citenamefont {Min},
  \citenamefont {Abedi}, \citenamefont {Kim},\ and\ \citenamefont
  {Gross}}]{Min:2014/prl/263004}%
  \BibitemOpen
  \bibfield  {author} {\bibinfo {author} {\bibfnamefont {S.~K.}\ \bibnamefont
  {Min}}, \bibinfo {author} {\bibfnamefont {A.}~\bibnamefont {Abedi}}, \bibinfo
  {author} {\bibfnamefont {K.~S.}\ \bibnamefont {Kim}}, \ and\ \bibinfo
  {author} {\bibfnamefont {E.}~\bibnamefont {Gross}},\ }\href {\doibase
  10.1103/PhysRevLett.113.263004} {\bibfield  {journal} {\bibinfo  {journal}
  {Phys. Rev. Lett.}\ }\textbf {\bibinfo {volume} {113}},\ \bibinfo {pages}
  {263004} (\bibinfo {year} {2014})}\BibitemShut {NoStop}%
\bibitem [{\citenamefont {K{\"o}uppel}, \citenamefont {Domcke},\ and\
  \citenamefont {Cederbaum}(1984)}]{Kouppel:1984/acp/59}%
  \BibitemOpen
  \bibfield  {author} {\bibinfo {author} {\bibfnamefont {H.}~\bibnamefont
  {K{\"o}uppel}}, \bibinfo {author} {\bibfnamefont {W.}~\bibnamefont {Domcke}},
  \ and\ \bibinfo {author} {\bibfnamefont {L.}~\bibnamefont {Cederbaum}},\
  }\href@noop {} {\bibfield  {journal} {\bibinfo  {journal} {Adv. Chem. Phys.}\
  }\textbf {\bibinfo {volume} {57}},\ \bibinfo {pages} {59} (\bibinfo {year}
  {1984})}\BibitemShut {NoStop}%
\bibitem [{\citenamefont {Ben-Nun}\ and\ \citenamefont
  {Martinez}(2002)}]{BenNun:2002tx}%
  \BibitemOpen
  \bibfield  {author} {\bibinfo {author} {\bibfnamefont {M.}~\bibnamefont
  {Ben-Nun}}\ and\ \bibinfo {author} {\bibfnamefont {T.~J.}\ \bibnamefont
  {Martinez}},\ }\href {\doibase 10.1002/0471264318.ch7} {\bibfield  {journal}
  {\bibinfo  {journal} {Adv. Chem. Phys.}\ }\textbf {\bibinfo {volume} {121}},\
  \bibinfo {pages} {439} (\bibinfo {year} {2002})}\BibitemShut {NoStop}%
\bibitem [{\citenamefont {Makhov}\ \emph {et~al.}(2017)\citenamefont {Makhov},
  \citenamefont {Symonds}, \citenamefont {Fernandez-Alberti},\ and\
  \citenamefont {Shalashilin}}]{Makhov:2017/cc/200}%
  \BibitemOpen
  \bibfield  {author} {\bibinfo {author} {\bibfnamefont {D.~V.}\ \bibnamefont
  {Makhov}}, \bibinfo {author} {\bibfnamefont {C.}~\bibnamefont {Symonds}},
  \bibinfo {author} {\bibfnamefont {S.}~\bibnamefont {Fernandez-Alberti}}, \
  and\ \bibinfo {author} {\bibfnamefont {D.~V.}\ \bibnamefont {Shalashilin}},\
  }\href {\doibase 10.1016/j.chemphys.2017.04.003} {\bibfield  {journal}
  {\bibinfo  {journal} {Chem. Phys.}\ }\textbf {\bibinfo {volume} {493}},\
  \bibinfo {pages} {200} (\bibinfo {year} {2017})}\BibitemShut {NoStop}%
\bibitem [{\citenamefont {Burghardt}, \citenamefont {Giri},\ and\ \citenamefont
  {Worth}(2008)}]{Burghardt:2008iz}%
  \BibitemOpen
  \bibfield  {author} {\bibinfo {author} {\bibfnamefont {I.}~\bibnamefont
  {Burghardt}}, \bibinfo {author} {\bibfnamefont {K.}~\bibnamefont {Giri}}, \
  and\ \bibinfo {author} {\bibfnamefont {G.~A.}\ \bibnamefont {Worth}},\ }\href
  {\doibase 10.1063/1.2996349} {\bibfield  {journal} {\bibinfo  {journal} {J.
  Chem. Phys.}\ }\textbf {\bibinfo {volume} {129}},\ \bibinfo {pages} {174104}
  (\bibinfo {year} {2008})}\BibitemShut {NoStop}%
\bibitem [{\citenamefont {Worth}, \citenamefont {Robb},\ and\ \citenamefont
  {Lasorne}(2008)}]{Worth:2008/MP/2077}%
  \BibitemOpen
  \bibfield  {author} {\bibinfo {author} {\bibfnamefont {G.~A.}\ \bibnamefont
  {Worth}}, \bibinfo {author} {\bibfnamefont {M.~A.}\ \bibnamefont {Robb}}, \
  and\ \bibinfo {author} {\bibfnamefont {B.}~\bibnamefont {Lasorne}},\ }\href
  {\doibase 10.1080/00268970802172503} {\bibfield  {journal} {\bibinfo
  {journal} {Mol. Phys.}\ }\textbf {\bibinfo {volume} {106}},\ \bibinfo {pages}
  {2077} (\bibinfo {year} {2008})}\BibitemShut {NoStop}%
\bibitem [{\citenamefont {Kan}(1981)}]{Kan:1981/pra/2831}%
  \BibitemOpen
  \bibfield  {author} {\bibinfo {author} {\bibfnamefont {K.-K.}\ \bibnamefont
  {Kan}},\ }\href {\doibase 10.1103/PhysRevA.24.2831} {\bibfield  {journal}
  {\bibinfo  {journal} {Phys. Rev. A}\ }\textbf {\bibinfo {volume} {24}},\
  \bibinfo {pages} {2831} (\bibinfo {year} {1981})}\BibitemShut {NoStop}%
\bibitem [{\citenamefont {Habershon}(2012)}]{Habershon:2012/jcp/014109}%
  \BibitemOpen
  \bibfield  {author} {\bibinfo {author} {\bibfnamefont {S.}~\bibnamefont
  {Habershon}},\ }\href {\doibase 10.1063/1.3671978} {\bibfield  {journal}
  {\bibinfo  {journal} {J. Chem. Phys.}\ }\textbf {\bibinfo {volume} {136}},\
  \bibinfo {pages} {014109} (\bibinfo {year} {2012})}\BibitemShut {NoStop}%
\bibitem [{\citenamefont {Thompson}, \citenamefont {Punwong},\ and\
  \citenamefont {Mart\'inez}(2010)}]{Thompson:2010/cp/70}%
  \BibitemOpen
  \bibfield  {author} {\bibinfo {author} {\bibfnamefont {A.~L.}\ \bibnamefont
  {Thompson}}, \bibinfo {author} {\bibfnamefont {C.}~\bibnamefont {Punwong}}, \
  and\ \bibinfo {author} {\bibfnamefont {T.~J.}\ \bibnamefont {Mart\'inez}},\
  }\href {\doibase 10.1016/j.chemphys.2010.03.020} {\bibfield  {journal}
  {\bibinfo  {journal} {Chem. Phys.}\ }\textbf {\bibinfo {volume} {370}},\
  \bibinfo {pages} {70} (\bibinfo {year} {2010})}\BibitemShut {NoStop}%
\bibitem [{\citenamefont {Izmaylov}(2013)}]{Izmaylov:2013/jcp/104115}%
  \BibitemOpen
  \bibfield  {author} {\bibinfo {author} {\bibfnamefont {A.~F.}\ \bibnamefont
  {Izmaylov}},\ }\href {\doibase 10.1063/1.4794047} {\bibfield  {journal}
  {\bibinfo  {journal} {J. Chem. Phys.}\ }\textbf {\bibinfo {volume} {138}},\
  \bibinfo {pages} {104115} (\bibinfo {year} {2013})}\BibitemShut {NoStop}%
\bibitem [{\citenamefont {Kramer}\ and\ \citenamefont
  {Saraceno}(1981)}]{Book/Kramer:1981}%
  \BibitemOpen
  \bibfield  {author} {\bibinfo {author} {\bibfnamefont {P.}~\bibnamefont
  {Kramer}}\ and\ \bibinfo {author} {\bibfnamefont {M.}~\bibnamefont
  {Saraceno}},\ }\href@noop {} {\emph {\bibinfo {title} {Geometry of the
  Time-Dependent Variational Principle in Quantum Mechanics}}}\ (\bibinfo
  {publisher} {Springer},\ \bibinfo {address} {New York},\ \bibinfo {year}
  {1981})\BibitemShut {NoStop}%
\bibitem [{\citenamefont {Broeckhove}\ \emph {et~al.}(1988)\citenamefont
  {Broeckhove}, \citenamefont {Lathouwers}, \citenamefont {Kesteloot},\ and\
  \citenamefont {Van~Leuven}}]{Broeckhove:1988wo}%
  \BibitemOpen
  \bibfield  {author} {\bibinfo {author} {\bibfnamefont {J.}~\bibnamefont
  {Broeckhove}}, \bibinfo {author} {\bibfnamefont {L.}~\bibnamefont
  {Lathouwers}}, \bibinfo {author} {\bibfnamefont {E.}~\bibnamefont
  {Kesteloot}}, \ and\ \bibinfo {author} {\bibfnamefont {P.}~\bibnamefont
  {Van~Leuven}},\ }\href@noop {} {\bibfield  {journal} {\bibinfo  {journal}
  {Chemical Physics Letters}\ }\textbf {\bibinfo {volume} {149}},\ \bibinfo
  {pages} {547} (\bibinfo {year} {1988})}\BibitemShut {NoStop}%
\bibitem [{\citenamefont {Burghardt}, \citenamefont {Meyer},\ and\
  \citenamefont {Cederbaum}(1999)}]{Burghardt:1999/jcp/2927}%
  \BibitemOpen
  \bibfield  {author} {\bibinfo {author} {\bibfnamefont {I.}~\bibnamefont
  {Burghardt}}, \bibinfo {author} {\bibfnamefont {H.-D.}\ \bibnamefont
  {Meyer}}, \ and\ \bibinfo {author} {\bibfnamefont {L.~S.}\ \bibnamefont
  {Cederbaum}},\ }\href {\doibase 10.1063/1.479574} {\bibfield  {journal}
  {\bibinfo  {journal} {J. Chem. Phys.}\ }\textbf {\bibinfo {volume} {111}},\
  \bibinfo {pages} {2927} (\bibinfo {year} {1999})}\BibitemShut {NoStop}%
\bibitem [{\citenamefont {Worth}\ and\ \citenamefont
  {Burghardt}(2003)}]{Worth:2003/cpl/502}%
  \BibitemOpen
  \bibfield  {author} {\bibinfo {author} {\bibfnamefont {G.~A.}\ \bibnamefont
  {Worth}}\ and\ \bibinfo {author} {\bibfnamefont {I.}~\bibnamefont
  {Burghardt}},\ }\href {\doibase 10.1016/S0009-2614(02)01920-6} {\bibfield
  {journal} {\bibinfo  {journal} {Chem. Phys. Lett.}\ }\textbf {\bibinfo
  {volume} {368}},\ \bibinfo {pages} {502} (\bibinfo {year}
  {2003})}\BibitemShut {NoStop}%
\bibitem [{\citenamefont {Richings}\ \emph {et~al.}(2015)\citenamefont
  {Richings}, \citenamefont {Polyak}, \citenamefont {Spinlove}, \citenamefont
  {Worth}, \citenamefont {Burghardt},\ and\ \citenamefont
  {Lasorne}}]{Richings:2015/irpc/269}%
  \BibitemOpen
  \bibfield  {author} {\bibinfo {author} {\bibfnamefont {G.~W.}\ \bibnamefont
  {Richings}}, \bibinfo {author} {\bibfnamefont {I.}~\bibnamefont {Polyak}},
  \bibinfo {author} {\bibfnamefont {K.~E.}\ \bibnamefont {Spinlove}}, \bibinfo
  {author} {\bibfnamefont {G.~A.}\ \bibnamefont {Worth}}, \bibinfo {author}
  {\bibfnamefont {I.}~\bibnamefont {Burghardt}}, \ and\ \bibinfo {author}
  {\bibfnamefont {B.}~\bibnamefont {Lasorne}},\ }\href {\doibase
  10.1080/0144235X.2015.1051354} {\bibfield  {journal} {\bibinfo  {journal}
  {Int. Rev. Phys. Chem.}\ }\textbf {\bibinfo {volume} {34}},\ \bibinfo {pages}
  {269} (\bibinfo {year} {2015})}\BibitemShut {NoStop}%
\bibitem [{\citenamefont {Plasser}\ \emph {et~al.}(2016)\citenamefont
  {Plasser}, \citenamefont {Ruckenbauer}, \citenamefont {Mai}, \citenamefont
  {Oppel}, \citenamefont {Marquetand},\ and\ \citenamefont
  {Gonz\'alez}}]{Plasser:2016/jctc/1207}%
  \BibitemOpen
  \bibfield  {author} {\bibinfo {author} {\bibfnamefont {F.}~\bibnamefont
  {Plasser}}, \bibinfo {author} {\bibfnamefont {M.}~\bibnamefont
  {Ruckenbauer}}, \bibinfo {author} {\bibfnamefont {S.}~\bibnamefont {Mai}},
  \bibinfo {author} {\bibfnamefont {M.}~\bibnamefont {Oppel}}, \bibinfo
  {author} {\bibfnamefont {P.}~\bibnamefont {Marquetand}}, \ and\ \bibinfo
  {author} {\bibfnamefont {L.}~\bibnamefont {Gonz\'alez}},\ }\href {\doibase
  10.1021/acs.jctc.5b01148} {\bibfield  {journal} {\bibinfo  {journal} {J.
  Chem. Theory Comput.}\ }\textbf {\bibinfo {volume} {12}},\ \bibinfo {pages}
  {1207} (\bibinfo {year} {2016})}\BibitemShut {NoStop}%
\bibitem [{\citenamefont {Malmqvist}(1986)}]{Malmqvist:1986/ijqc/479}%
  \BibitemOpen
  \bibfield  {author} {\bibinfo {author} {\bibfnamefont {P.~{\AA}.}\
  \bibnamefont {Malmqvist}},\ }\href {\doibase 10.1002/qua.560300404}
  {\bibfield  {journal} {\bibinfo  {journal} {Int. J. Quantum Chem.}\ }\textbf
  {\bibinfo {volume} {30}},\ \bibinfo {pages} {479} (\bibinfo {year}
  {1986})}\BibitemShut {NoStop}%
\bibitem [{\citenamefont {Osamura}(1989)}]{Osamura:1989/tca/113}%
  \BibitemOpen
  \bibfield  {author} {\bibinfo {author} {\bibfnamefont {Y.}~\bibnamefont
  {Osamura}},\ }\href {\doibase 10.1007/BF00532128} {\bibfield  {journal}
  {\bibinfo  {journal} {Theor. Chim. Acta}\ }\textbf {\bibinfo {volume} {76}},\
  \bibinfo {pages} {113} (\bibinfo {year} {1989})}\BibitemShut {NoStop}%
\bibitem [{\citenamefont {Handy}\ and\ \citenamefont
  {Schaefer}(1984)}]{Handy:1984/jcp/5031}%
  \BibitemOpen
  \bibfield  {author} {\bibinfo {author} {\bibfnamefont {N.~C.}\ \bibnamefont
  {Handy}}\ and\ \bibinfo {author} {\bibfnamefont {H.~F.}\ \bibnamefont
  {Schaefer}},\ }\href {\doibase 10.1063/1.447489} {\bibfield  {journal}
  {\bibinfo  {journal} {J. Chem. Phys.}\ }\textbf {\bibinfo {volume} {81}},\
  \bibinfo {pages} {5031} (\bibinfo {year} {1984})}\BibitemShut {NoStop}%
\bibitem [{\citenamefont {Lengsfield}, \citenamefont {Saxe},\ and\
  \citenamefont {Yarkony}(1984)}]{Lengsfield:1984/jcp/4549}%
  \BibitemOpen
  \bibfield  {author} {\bibinfo {author} {\bibfnamefont {B.~H.}\ \bibnamefont
  {Lengsfield}}, \bibinfo {author} {\bibfnamefont {P.}~\bibnamefont {Saxe}}, \
  and\ \bibinfo {author} {\bibfnamefont {D.~R.}\ \bibnamefont {Yarkony}},\
  }\href {\doibase 10.1063/1.447428} {\bibfield  {journal} {\bibinfo  {journal}
  {J. Chem. Phys.}\ }\textbf {\bibinfo {volume} {81}},\ \bibinfo {pages} {4549}
  (\bibinfo {year} {1984})}\BibitemShut {NoStop}%
\bibitem [{\citenamefont {Gill}, \citenamefont {Johnson},\ and\ \citenamefont
  {Pople}(1994)}]{Gill:1994/cpl/65}%
  \BibitemOpen
  \bibfield  {author} {\bibinfo {author} {\bibfnamefont {P.~M.~W.}\
  \bibnamefont {Gill}}, \bibinfo {author} {\bibfnamefont {B.~G.}\ \bibnamefont
  {Johnson}}, \ and\ \bibinfo {author} {\bibfnamefont {J.~A.}\ \bibnamefont
  {Pople}},\ }\href {\doibase 10.1016/0009-2614(93)E1340-M} {\bibfield
  {journal} {\bibinfo  {journal} {Chem. Phys. Lett.}\ }\textbf {\bibinfo
  {volume} {217}},\ \bibinfo {pages} {65} (\bibinfo {year} {1994})}\BibitemShut
  {NoStop}%
\bibitem [{\citenamefont {Shin}\ and\ \citenamefont
  {Metiu}(1995)}]{Shin:1995/jcp/9285}%
  \BibitemOpen
  \bibfield  {author} {\bibinfo {author} {\bibfnamefont {S.}~\bibnamefont
  {Shin}}\ and\ \bibinfo {author} {\bibfnamefont {H.}~\bibnamefont {Metiu}},\
  }\href {\doibase 10.1063/1.468795} {\bibfield  {journal} {\bibinfo  {journal}
  {J. Chem. Phys.}\ }\textbf {\bibinfo {volume} {102}},\ \bibinfo {pages}
  {9285} (\bibinfo {year} {1995})}\BibitemShut {NoStop}%
\bibitem [{\citenamefont {Shin}\ and\ \citenamefont
  {Metiu}(1996)}]{Shin:1996/jpc/7867}%
  \BibitemOpen
  \bibfield  {author} {\bibinfo {author} {\bibfnamefont {S.}~\bibnamefont
  {Shin}}\ and\ \bibinfo {author} {\bibfnamefont {H.}~\bibnamefont {Metiu}},\
  }\href {\doibase 10.1021/jp952498a} {\bibfield  {journal} {\bibinfo
  {journal} {J. Phys. Chem.}\ }\textbf {\bibinfo {volume} {100}},\ \bibinfo
  {pages} {7867} (\bibinfo {year} {1996})}\BibitemShut {NoStop}%
\bibitem [{\citenamefont {Hader}\ \emph {et~al.}(2017)\citenamefont {Hader},
  \citenamefont {Albert}, \citenamefont {Gross},\ and\ \citenamefont
  {Engel}}]{Hader:2017/jcp/074304}%
  \BibitemOpen
  \bibfield  {author} {\bibinfo {author} {\bibfnamefont {K.}~\bibnamefont
  {Hader}}, \bibinfo {author} {\bibfnamefont {J.}~\bibnamefont {Albert}},
  \bibinfo {author} {\bibfnamefont {E.~K.~U.}\ \bibnamefont {Gross}}, \ and\
  \bibinfo {author} {\bibfnamefont {V.}~\bibnamefont {Engel}},\ }\href
  {\doibase 10.1063/1.4975811} {\bibfield  {journal} {\bibinfo  {journal} {J.
  Chem. Phys.}\ }\textbf {\bibinfo {volume} {146}},\ \bibinfo {pages} {074304}
  (\bibinfo {year} {2017})}\BibitemShut {NoStop}%
\bibitem [{Mat(2012)}]{Matlab:2012b}%
  \BibitemOpen
  \href@noop {} {}\bibinfo {howpublished} {MATLAB, version 8.0.0.783 (R2012b).
  The MathWorks Inc., Natick, Massachusetts} (\bibinfo {year}
  {2012})\BibitemShut {NoStop}%
\bibitem [{\citenamefont {Sch\"{o}n}\ and\ \citenamefont
  {K\"{o}ppel}(1995)}]{Schon:1995/jcp/9292}%
  \BibitemOpen
  \bibfield  {author} {\bibinfo {author} {\bibfnamefont {J.}~\bibnamefont
  {Sch\"{o}n}}\ and\ \bibinfo {author} {\bibfnamefont {H.}~\bibnamefont
  {K\"{o}ppel}},\ }\href {\doibase 10.1063/1.469988} {\bibfield  {journal}
  {\bibinfo  {journal} {J. Chem. Phys.}\ }\textbf {\bibinfo {volume} {103}},\
  \bibinfo {pages} {9292} (\bibinfo {year} {1995})}\BibitemShut {NoStop}%
\bibitem [{\citenamefont {Zwanziger}\ and\ \citenamefont
  {Grant}(1987)}]{Zwanziger:1987/jcp/2954}%
  \BibitemOpen
  \bibfield  {author} {\bibinfo {author} {\bibfnamefont {J.~W.}\ \bibnamefont
  {Zwanziger}}\ and\ \bibinfo {author} {\bibfnamefont {E.~R.}\ \bibnamefont
  {Grant}},\ }\href {\doibase 10.1063/1.453083} {\bibfield  {journal} {\bibinfo
   {journal} {J. Chem. Phys.}\ }\textbf {\bibinfo {volume} {87}},\ \bibinfo
  {pages} {2954} (\bibinfo {year} {1987})}\BibitemShut {NoStop}%
\bibitem [{\citenamefont {Domcke}\ and\ \citenamefont
  {Yarkony}(2012)}]{Domcke:2012/arpc/325}%
  \BibitemOpen
  \bibfield  {author} {\bibinfo {author} {\bibfnamefont {W.}~\bibnamefont
  {Domcke}}\ and\ \bibinfo {author} {\bibfnamefont {D.~R.}\ \bibnamefont
  {Yarkony}},\ }\href {\doibase 10.1146/annurev-physchem-032210-103522}
  {\bibfield  {journal} {\bibinfo  {journal} {Annu. Rev. Phys. Chem.}\ }\textbf
  {\bibinfo {volume} {63}},\ \bibinfo {pages} {325} (\bibinfo {year}
  {2012})}\BibitemShut {NoStop}%
\bibitem [{\citenamefont {Izmaylov}\ and\ \citenamefont
  {Joubert-Doriol}(2017)}]{Izmaylov:2017/jpcl/1793}%
  \BibitemOpen
  \bibfield  {author} {\bibinfo {author} {\bibfnamefont {A.~F.}\ \bibnamefont
  {Izmaylov}}\ and\ \bibinfo {author} {\bibfnamefont {L.}~\bibnamefont
  {Joubert-Doriol}},\ }\href {\doibase 10.1021/acs.jpclett.7b00596} {\bibfield
  {journal} {\bibinfo  {journal} {J. Phys. Chem. Lett.}\ }\textbf {\bibinfo
  {volume} {8}},\ \bibinfo {pages} {1793} (\bibinfo {year} {2017})}\BibitemShut
  {NoStop}%
\bibitem [{\citenamefont {R\"omer}\ and\ \citenamefont
  {Burghardt}(2013)}]{Romer:2013/mp/3618}%
  \BibitemOpen
  \bibfield  {author} {\bibinfo {author} {\bibfnamefont {S.}~\bibnamefont
  {R\"omer}}\ and\ \bibinfo {author} {\bibfnamefont {I.}~\bibnamefont
  {Burghardt}},\ }\href {\doibase 10.1080/00268976.2013.844371} {\bibfield
  {journal} {\bibinfo  {journal} {Mol. Phys.}\ }\textbf {\bibinfo {volume}
  {111}},\ \bibinfo {pages} {3618} (\bibinfo {year} {2013})}\BibitemShut
  {NoStop}%
\bibitem [{\citenamefont {Joubert-Doriol}\ and\ \citenamefont
  {Izmaylov}(2015)}]{Joubert:2015/jcp/134107}%
  \BibitemOpen
  \bibfield  {author} {\bibinfo {author} {\bibfnamefont {L.}~\bibnamefont
  {Joubert-Doriol}}\ and\ \bibinfo {author} {\bibfnamefont {A.~F.}\
  \bibnamefont {Izmaylov}},\ }\href {\doibase 10.1063/1.4916384} {\bibfield
  {journal} {\bibinfo  {journal} {J. Chem. Phys.}\ }\textbf {\bibinfo {volume}
  {142}},\ \bibinfo {pages} {134107} (\bibinfo {year} {2015})}\BibitemShut
  {NoStop}%
\bibitem [{\citenamefont {Vacher}\ \emph {et~al.}(2017)\citenamefont {Vacher},
  \citenamefont {Bearpark}, \citenamefont {Robb},\ and\ \citenamefont
  {Malhado}}]{Vacher:2017/prl/083001}%
  \BibitemOpen
  \bibfield  {author} {\bibinfo {author} {\bibfnamefont {M.}~\bibnamefont
  {Vacher}}, \bibinfo {author} {\bibfnamefont {M.~J.}\ \bibnamefont
  {Bearpark}}, \bibinfo {author} {\bibfnamefont {M.~A.}\ \bibnamefont {Robb}},
  \ and\ \bibinfo {author} {\bibfnamefont {J.~P.}\ \bibnamefont {Malhado}},\
  }\href {\doibase 10.1103/PhysRevLett.118.083001} {\bibfield  {journal}
  {\bibinfo  {journal} {Phys. Rev. Lett.}\ }\textbf {\bibinfo {volume} {118}},\
  \bibinfo {pages} {083001} (\bibinfo {year} {2017})}\BibitemShut {NoStop}%
\end{thebibliography}
%

\end{document}